\documentclass[%
 reprint,
 amsmath,amssymb,
 aps,
]{revtex4-2}

\usepackage{graphicx}
\usepackage{dcolumn}
\usepackage{bm}

\usepackage{multirow}
\usepackage{hyperref}
\usepackage{tikz}
\usetikzlibrary{arrows.meta, positioning, calc, shapes.arrows} 

\begin{document}

\preprint{APS/123-QED}

\title{Centre-of-momentum frame analysis of $\eta$ production in DUNE}

\author{R K Pradhan\textsuperscript{1}}%
 \email{kumarriteshpradhan@gmail.com}

\author{R Lalnuntluanga\textsuperscript{2}}
\email{tluangaralte.phy@gmail.com}

\author{A Giri\textsuperscript{1}}%
 \email{giria@phy.iith.ac.in}
 
\affiliation{%
\textsuperscript{1}
 \textit{Indian Institute of Technology Hyderabad, Hyderabad, 502284, Telangana, India}
}%

\affiliation{\textsuperscript{2}
\textit{Tel Aviv University, Tel Aviv 69978, Israel}
}


\begin{abstract}
A deep understanding of neutrino-nucleus interaction is crucial for the precise measurement of neutrino oscillation parameters and cross section measurements. Various nuclear effects, such as initial state effects (IS) and Final state interactions (FSI), make neutrino interactions more complicated. To probe the impacts of the nuclear effects, a separate study of FSI and IS is required. A set of variables known as Centre-of-momentum (c.m.) variables ($\theta_{c.m.}$ and $E_{c.m.}$) provides a unique approach to isolate the FSI effect with minimal sensitivity to IS. This work presents the importance of c.m. variables in neutrino-induced eta ($\eta$) meson production in the DUNE near detector. $\theta_{c.m.}$ is an important parameter to improve the FSI modeling, while $E_{c.m.}$ helps in isolating high-purity neutrino-Hydrogen events. The study of $\eta$ production in neutrino interactions helps in understanding the theoretical descriptions of higher resonance states.

\end{abstract}

\maketitle


\section{Introduction}

Accurate modeling of neutrino-nucleus interactions is crucial for the precise measurement of neutrino oscillation parameters, including the leptonic CP-violating phase ($\delta_{CP}$) measurement. The upcoming long-baseline neutrino experiments such as the Deep Underground Neutrino Experiment (DUNE) \cite{DUNE:2020ypp} and Hyper-Kamiokande \cite{Hyper-Kamiokande:2018ofw} provide a broad physics program that focuses on a deep understanding of neutrino interaction at the nuclear level, to enhance the upcoming BSM physics programs \cite{Balantekin:2022jrq}. These experiments use GeV-scale neutrino beams from the accelerator facility. Resonant interaction (RES) is an important process in the $\mathcal{O}$(GeV) energy range, where a neutrino excites a single nucleon to a higher baryonic state. The study of nucleon resonances provides valuable insight into the intrinsic structure of the nucleon. One of the major challenges in hadron physics is the discrepancy between the observed baryonic spectra and theoretical calculations regarding the number and properties of excited baryonic states \cite{Aznauryan:2009da}. Neutrinos can excite nucleons to a wide range of baryon resonances \cite{Rein:1980wg,Leitner:2008ue}, and the current studies show that the most dominant resonant state is $\Delta$(1232). However, the higher order RES still contribute $\sim$10\% to the overall event rate \cite{MicroBooNE:2023ubu}. If these contributions are not accurately modeled, they can result in background mismodeling, affecting the precision of neutrino oscillation measurements and searches for BSM physics. The effect of the second resonance region ($P_{11}$(1440), $D_{13}$(1520), and $S_{11}$ (1535)) on neutrino interaction is studied in Ref. \cite{Lalnuntluanga:2023qkp}. RES interactions are one of the dominant channels in DUNE, and uncertainties in their modeling contribute significantly to the total systematics. 

Most of the RES states decay to a pion and a nucleon; however, they can also produce other mesons such as $\eta,\,K,\,\rho$, etc. $\eta$ can be produced from RES such as N(1535), N (1650), and N(1710) with branching fractions of 30\%-55\%, 15\%-35\%, and 10\%-50\%, respectively \cite{MicroBooNE:2023ubu}.  The study of $\eta$ production in neutrino interaction provides a unique way to explore higher RES states, which is not easy to study through pion production. In addition, measurement of $\eta$ is also important in BSM physics searches, such as observation of $\eta$ opens a way to search proton decay via $p \rightarrow e^+\,\eta$ and $p \rightarrow \mu^+\, \eta$ channels in DUNE. The decay of $\eta$ through photon pairs offers a high-energy electromagnetic (EM) shower sample, enabling calibration of EM energy scales in detectors like LArTPCs. This complements lower-energy $\pi^0$ calibrations \cite{MicroBooNE:2019rgx} and improves the accuracy of electron neutrino energy reconstruction, vital for measuring $\delta_{CP}$ and other oscillation parameters. The $\eta$ production has been measured by MicroBooNE on Argon target \cite{MicroBooNE:2023ubu} and on $Ne-H_2$ target by the BEBC WA59 Collaboration \cite{BEBCWA59:1989ofp}. High-statistics measurements of $\eta$ by upcoming neutrino experiments will further refine RES interaction models. 

\begin{figure*}
    \centering
    \includegraphics[width=19cm,height=5.5cm]{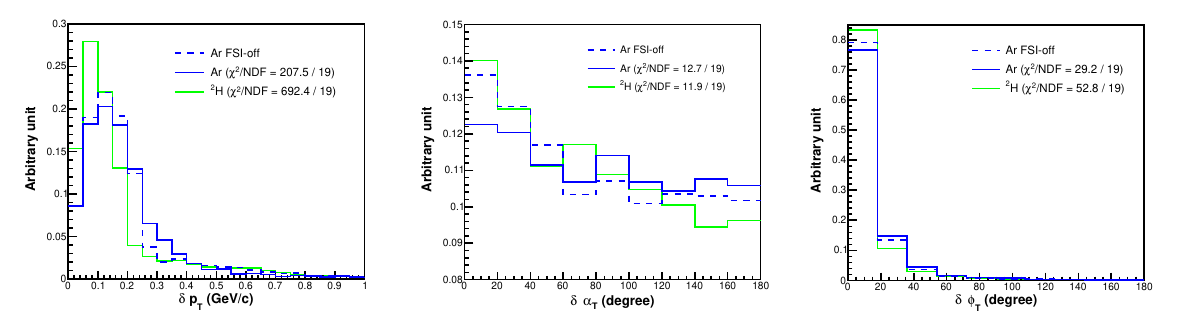}
    \caption{Area-normalized distribution of TKI variables for the 1p1$\eta$ channel on argon and deuterium ($\eta$ cannot be produced in $\nu$–H interactions in GENIE). The difference between Ar-FSI-off and Ar indicates the dependence on FSI, while the difference between Ar-FSI-off and $^2$H shows the IS dependence for $\delta p_T$ and $\delta \alpha_T$. However, $\delta \phi_T$ shows minimal dependence on both IS and FSI.}
    \label{tki_variables}
\end{figure*}

The complexity of the nuclear environment introduces various nuclear effects: the initial state of nucleon (IS) and final state interactions (FSI) \cite{Dytman:2009zz}. The FSI includes the rescattering of the hadrons produced inside the nuclear medium. Due to FSI, the event topology of a particular interaction can be misidentified. A RES event can be misidentified as non-RES due to the pion absorption, or a QE misidentified as non-QE due to pion production. Hence, a proper event selection is required to study a particular process \cite{Lalnuntluanga:2024lti}. Additionally, IS and FSI cannot be directly measured by current detectors, which makes them a significant cause of systematic uncertainties. The nuclear effects can be explored by using various Methods such as the transverse kinematic imbalance (TKI) \cite{Lu:2015tcr}, generalized kinematic imbalance (GKI) \cite{MicroBooNE:2023krv}, or longitudinal kinematic imbalance \cite{Baudis:2023tma}. The previous measurement of TKI variables has improved the theoretical descriptions of neutrino interactions in Monte-Carlo (MC) studies \cite{GENIE:2024ufm,Yan:2024kkg,Pradhan:2024gqv}; however, TKI can't measure a specific nuclear effect as it is sensitive to both IS and FSI \cite{Li:2025iiv}. The dependence of TKI variables on both FSI and IS is shown in Fig. \ref{tki_variables}. The longitudinal variable, $p_{long}$, is only sensitive to the removal energy. The new variables like center-of-momentum (c.m.) variables ($\theta_{c.m.},E_{c.m.}$) can measure the FSI minimally dependent on the IS \cite{Li:2025iiv}. The study of the c.m. variables for the single pion-single proton system in Ref.\cite{Li:2025iiv} shows that these variables are useful for advanced modeling of FSI.

In this work, we explore the c.m. variables for neutrino-induced $\eta$ production in the charge current (CC) neutrino-Argon interactions in DUNE near detector (ND). This work presents the MC analysis on the single proton - single eta system. It shows that the c.m. variable $\theta_{c.m.}$ is independent of IS, and is only sensitive to the FSI. The details of the analysis and simulation are explained in  Sec. \ref{formalism}. The results and conclusions are discussed in Secs. \ref{results} and \ref{conclusion}, respectively.

\section{Formalism \label{formalism}}
\subsection{C.M. Variables}

Consider the most contributing RES state N(1535) that decays to $\eta$ via \cite{ParticleDataGroup:2022pth}
\begin{equation*}
    N^+ \rightarrow \eta\,+\,p
\end{equation*}

\begin{figure}[!h]
    \centering
    \includegraphics[width=9cm,height=5.3cm]{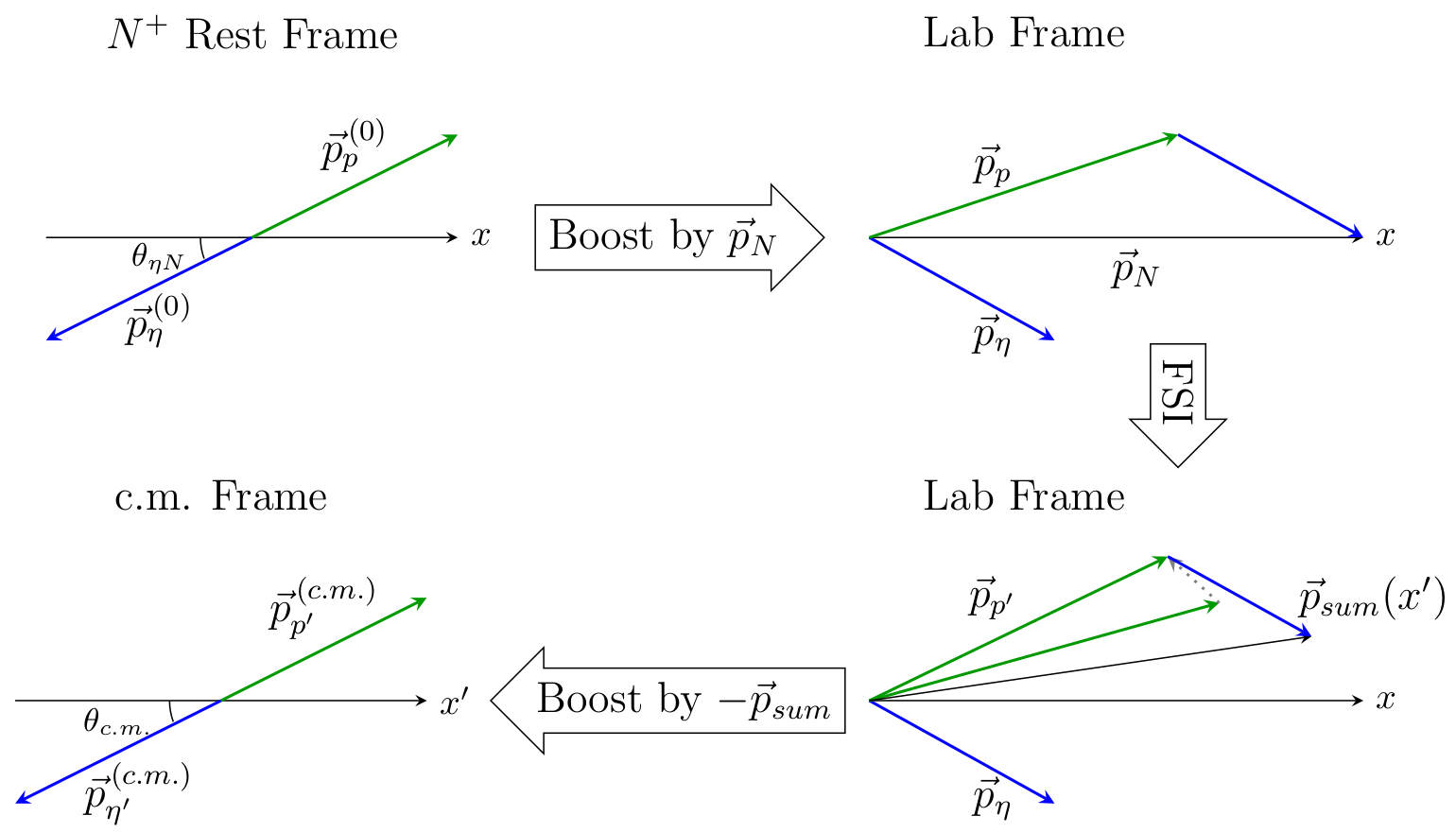}
    \caption{Schematic illustration of $N^+$ decay in its rest frame (without FSI), in lab frame, and c.m. frame (after FSI) \cite{Li:2025iiv}.}
    \label{fig:COM}
\end{figure}

In the rest frame of N(1535), it decays to proton and eta in a back-to-back configuration as shown in the top left in Fig. \ref{fig:COM}. $\theta_{\eta N}$ is the decay angle of $\eta$ in the $N^+$ rest frame; the angle between $\vec{p}_{\eta}^{(0)}$ and the $x$-axis. The $x$-axis is taken along the momentum of $N^+$ ($\vec{p}_N$) in the Lab frame. The kinematics of the proton and the eta in the Lab frame can be obtained by boosting the rest frame with $\vec{p}_N$ shown in the top right of Fig. \ref{fig:COM}. $\vec{p}_N$ is the sum of $\vec{p}_{\eta}$ and $\vec{p}_p$ without FSI. With the effect of FSI, the produced hadrons undergo multiple rescatterings. The propagation of $\eta$ inside the nucleus can be affected by FSI. However, $\eta$ decays outside the nucleus because of its lifetime of more than $10^{-19}$ seconds. The kinematics of the proton and eta are altered due to the FSI shown on the bottom right of Fig. \ref{fig:COM}. 
The sum of the altered momenta ($\vec{p}_{sum}$ = $\vec{p}_{\eta}\,'$ + $\vec{p}_p\,'$) differs from $\vec{p}_N$ due to FSI. The kinematics in the $N^+$ rest frame can be accessed by boosting the Lab frame to the c.m. frame with $-\vec{p}_{sum}$ shown on the bottom left of Fig. \ref{fig:COM}, taking the $x'$-axis along the $\vec{p}_{sum}$ in the Lab frame. The decay angle of $\eta$ ($\theta_{c.m.}$) in the c.m. frame, the angle between $\vec{p}_{\eta'}^{c.m.}$ and $x'$-axis, is different than $\theta_{\eta N}$ mostly due to FSI.

The total energy in the c.m. frame ($E_{c.m.}$) is another c.m. variable that helps to select the events with minimal FSI effects. Without the effect of FSI, $E_{c.m.}$ is equal to the mass of the RES state. A cut on $E_{c.m.}$ within the range of the RES mass can help in selecting the events with minimal FSI, similar to $\nu$-H events.

\subsection{Reconstruction of $\eta$}

$\eta$ mesons are neutral and unstable with life time of order $10^{-19}$ seconds. They can be detected indirectly from the decay products. The dominant decay modes are $\eta \rightarrow 2\gamma$, $\eta \rightarrow 3\pi^0$, and $\eta \rightarrow \pi^0\pi^+\pi^-$ with branching ratios of 40\%, 33\%, and 23\%, respectively \cite{ParticleDataGroup:2022pth}. This work considers the first two dominant decay channels to reconstruct $\eta$ event. The invariant mass of photon pairs is calculated using,
\begin{equation*}
    M_{\gamma \gamma} = \sqrt{2E_1E_2(1-\cos\theta)}
\end{equation*}
where, $E_1$ and $E_2$ are the energies of photon pair and $\theta$ is the angle between them. For the decay mode, $\eta \rightarrow 2\gamma$, a peak around 548 MeV/$c^2$ in the invariant mass distribution of photon pairs indicates an $\eta$ event. For the channel, $\eta \rightarrow 3\pi^0$, each $\pi^0$ can be reconstructed from a peak around  135 MeV/$c^2$ in the $M_{\gamma \gamma}$ distribution. An $\eta$ event can be identified from a peak around 548 MeV/$c^2$ in the invariant mass of the three $\pi^0$.

\subsection{Simulation Details}

GENIE v3.06.00 \cite{Andreopoulos:2015wxa} is used for simulating the $\nu_{\mu}$ CC interaction with Argon nucleus using DUNE-ND $\nu_{\mu}$ flux that peaks around 2.5 GeV \cite{DUNE:2021cuw}. The simulation considers all possible CC interactions, including quasielastic (QE), resonance (RES), coherent (COH), meson exchange (MEC), and deep inelastic scattering (DIS). Samples of 1 million $\nu_{\mu}$ events are simulated by enabling different nuclear effects and the nuclear models. The ground state of the nucleus is considered in the simulation using the Local Fermi Gas model (LFG) \cite{Bodek:1981wr}, Correlated Fermi Gas model (CFG) \cite{CLAS:2005ola}, Spectral Function (SF) \cite{Benhar:1994hw}, Spectral function-like-LFG, and Spectral function-like-CFG. The default removal energy of the nucleon for argon is taken as 28 MeV. The models used for different channels are: the Valencia QE model \cite{Gran:2013kda} for CCQE, the Valencia MEC model \cite{Nieves:2016sma} for CCMEC, the Berger-Sehgal (BS) model \cite{Berger:2007rq} for CCRES, and the Bodek-Yang model \cite{Bodek:2002vp} for CCDIS. All 17 resonances are included in the simulation. The RES-DIS joining scheme is considered with the invariant mass threshold of 1.9 GeV. The RES axial mass is set to its default value of 1.09 GeV. The INTRANUKE hA and hN models \cite{GENIE:2021npt} are used for FSI effects in the simulation. The events with the final state containing 1 proton and 1 $\eta$ are selected for the analysis. Due to FSI, there could be extra nucleons and pions in the final state along with 1p1$\eta$. There are also backgrounds from DIS in $\eta$ production.

\section{Results \label{results}}
\subsection{Decay angle of $\eta$ in c.m. frame ($\theta_{c.m.}$)}

\begin{figure}[!h]
    \centering
    \includegraphics[width=9.5cm,height=7cm]{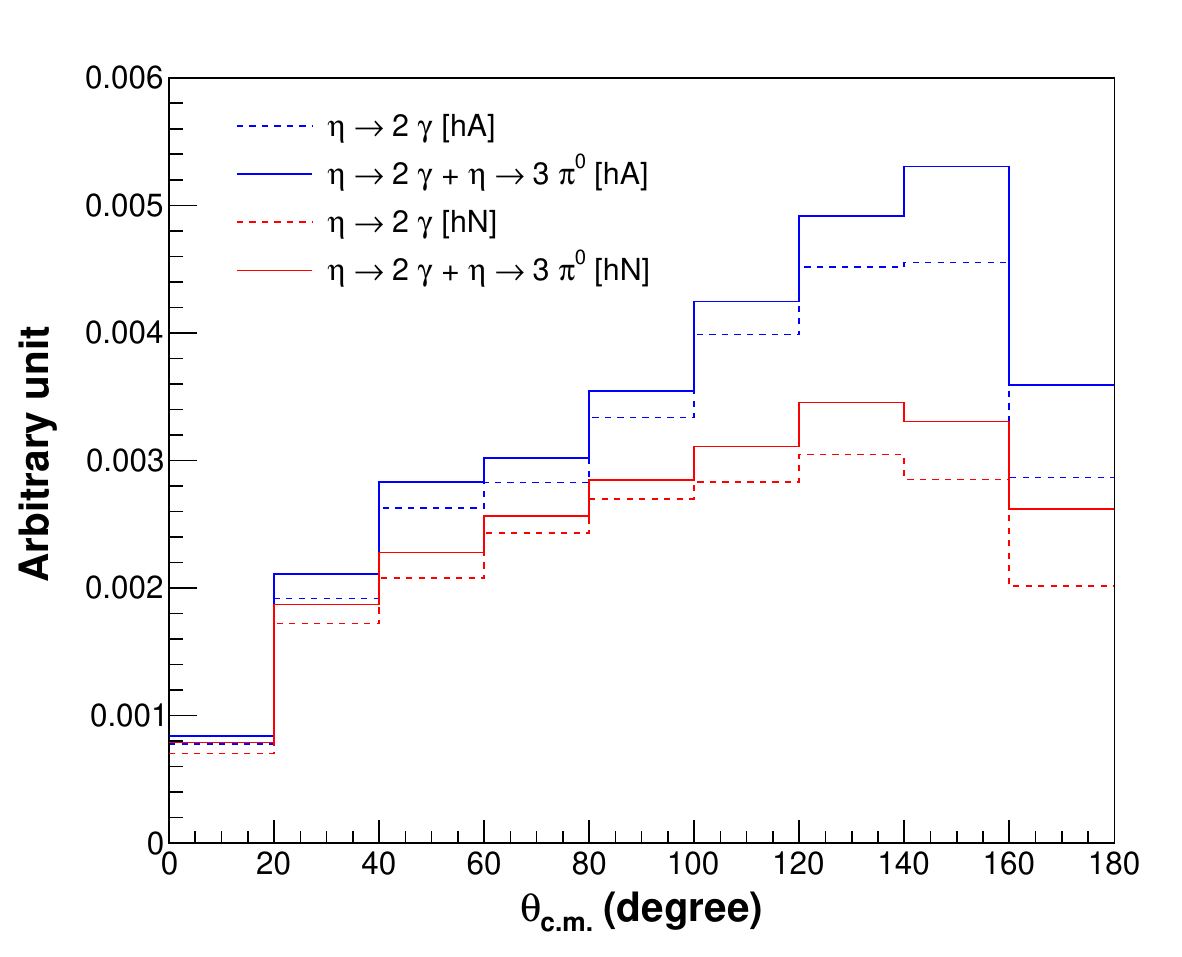}
    \caption{Cross section normalized $\theta_{c.m.}$ distribution for decay channel only $\eta \rightarrow 2\gamma$ (dotted) and both $\eta \rightarrow 2\gamma$, and $\eta \rightarrow 3\pi^0$ (solid) for hA (blue) and hN (Red) FSI models. }
    \label{fig:theta_decay}
\end{figure}

The decay angle of $\eta$ in the c.m. frame is reconstructed from the decay channels of $\eta \rightarrow 2\gamma$ and $\eta \rightarrow 3\pi^0$ using the FSI model hA and hN shown in Fig. \ref{fig:theta_decay}. There is an increment in the cross section when both channels are considered however, the shapes of the curves are the same for the channels. For further analysis, both channels are considered to reconstruct the $\eta$. The distribution of $\theta_{c.m.}$ differs in both shape and cross section for FSI model hA and hN, indicating a high dependence of $\theta_{c.m.}$ on the FSI models.

\begin{figure}[!h]
    \centering
    \includegraphics[width=9.5cm,height=7cm]{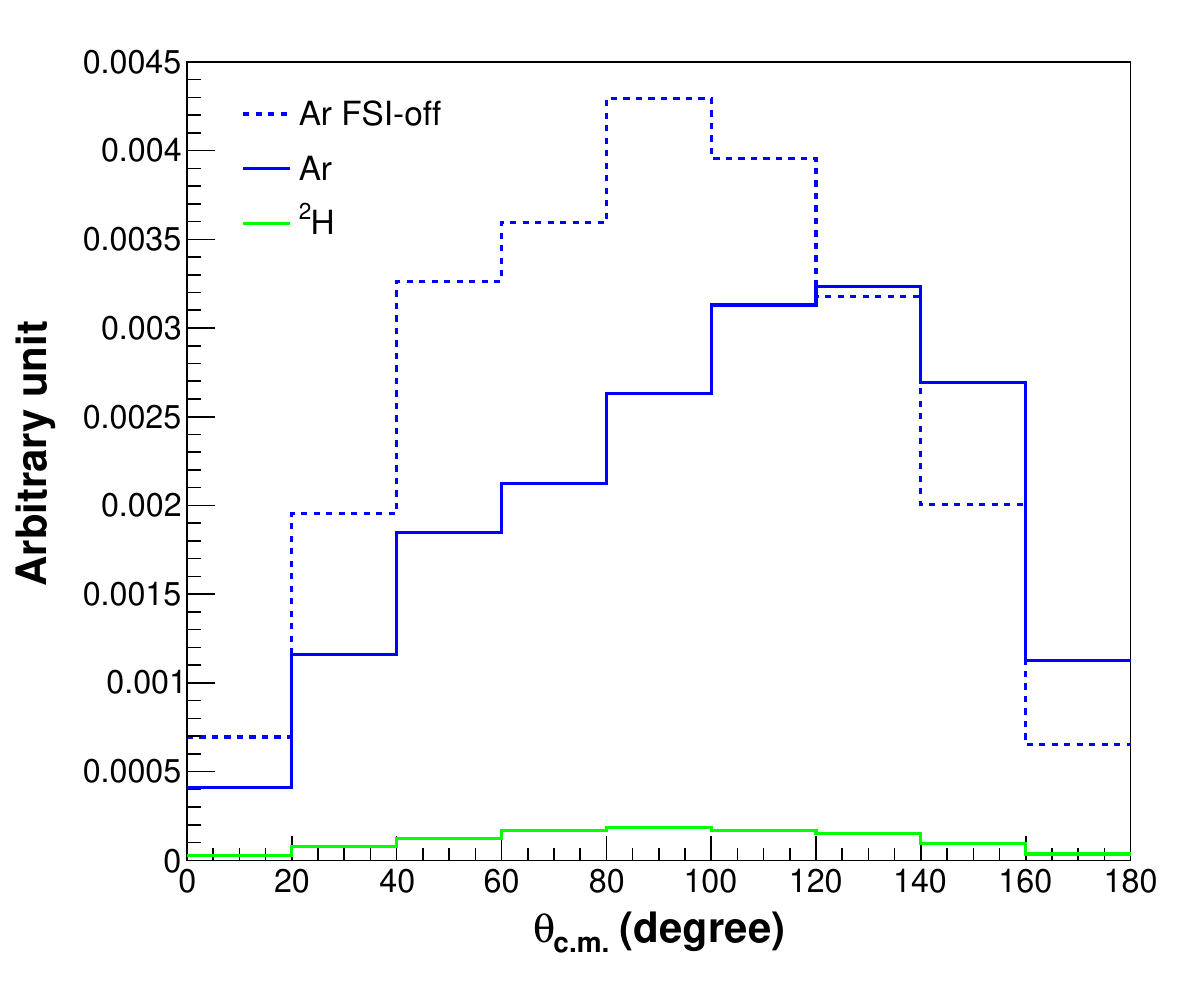}
    \caption{Cross section normalized distribution of $\theta_{c.m.}$ for Argon-FSI-ON/OFF and Deuterium.}
    \label{fig:1}
\end{figure}

\begin{figure*}
    \centering
    \includegraphics[width=19cm,height=7cm]{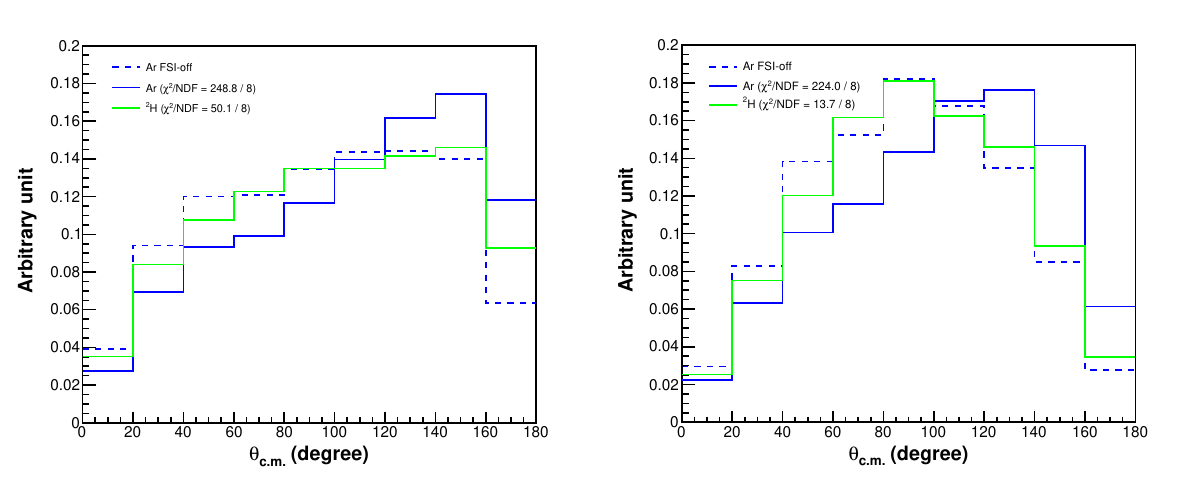}
    \caption{Area normalized distribution of $\theta_{c.m.}$ for Argon-FSI-ON/OFF and Deuterium considering $\nu_{\mu}$CC1p1$\eta$ events (left) and resonance events (right). }
    \label{fig:2}
\end{figure*}

To probe the effect of FSI, $\theta_{c.m.}$ is reconstructed with enabling and disabling FSI shown in Fig. \ref{fig:1}. There is a drastic change in the distribution when FSI is disabled. Without FSI, $\theta_{c.m.}$ is symmetric and peaks around 90$^\circ$, whereas it is shifted towards higher angles, peaking around 130$^\circ$ when FSI is enabled. This shows a strong $\theta_{c.m.}$ dependence on FSI. The impact of IS can be isolated by comparing the shape of the distribution for FSI-off with $\nu$-H events. Since Hydrogen is only one proton, $\nu$-H interactions won't be affected by FSI. Hence, an overlap between FSI-off and $\nu$-H curves confirms the insensitivity of $\theta_{c.m.}$ to the IS. However, $\eta$ can't be produced in $\nu$-H interactions in GENIE as a neutrino can't excite the Hydrogen nucleus to a resonance state higher than $\Delta$(1232). To achieve this, deuterium ($^2$H) target is used. Since it contains one proton and one neutron, the FSI effects are assumed to be minimal. For shape comparison, the $\chi^2$ per degree of freedom (NDF) is calculated for each shape-comparison plot with respect to the first curve in the respective plot. The shape-comparison between $\nu$-Ar FSI off and $\nu$-$^2$H is shown in Fig. \ref{fig:2}. The left panel shows the area-normalized distribution of $\theta_{c.m.}$ for $\nu_{\mu}$CC1p1$\eta$ events. The Ar-FSI-off distribution differs from $^2$H with a $\chi^2$/NDF of 50.1/8. If only events from RES channels are considered, shown in the right panel, the $\chi^2$/NDF is reduced to 13.7/8, and the Ar-FSI-off distribution is close to the $^2$H. The large $\chi^2$/NDF in the selection could be because of the contributions from DIS and minimal FSI effects in the deuterium nucleus. A considerably high $\chi^2$/NDF in the case of Ar-FSI-on, both for the selection and the RES events (248.8/8 and 224.0/8, respectively), shows the strong impact of FSI than IS on $\theta_{c.m.}$.

\subsection{Selection of no-FSI events using $E_{c.m.}$}

\begin{figure}[!h]
    \centering
    \includegraphics[width=9.5cm,height=7cm]{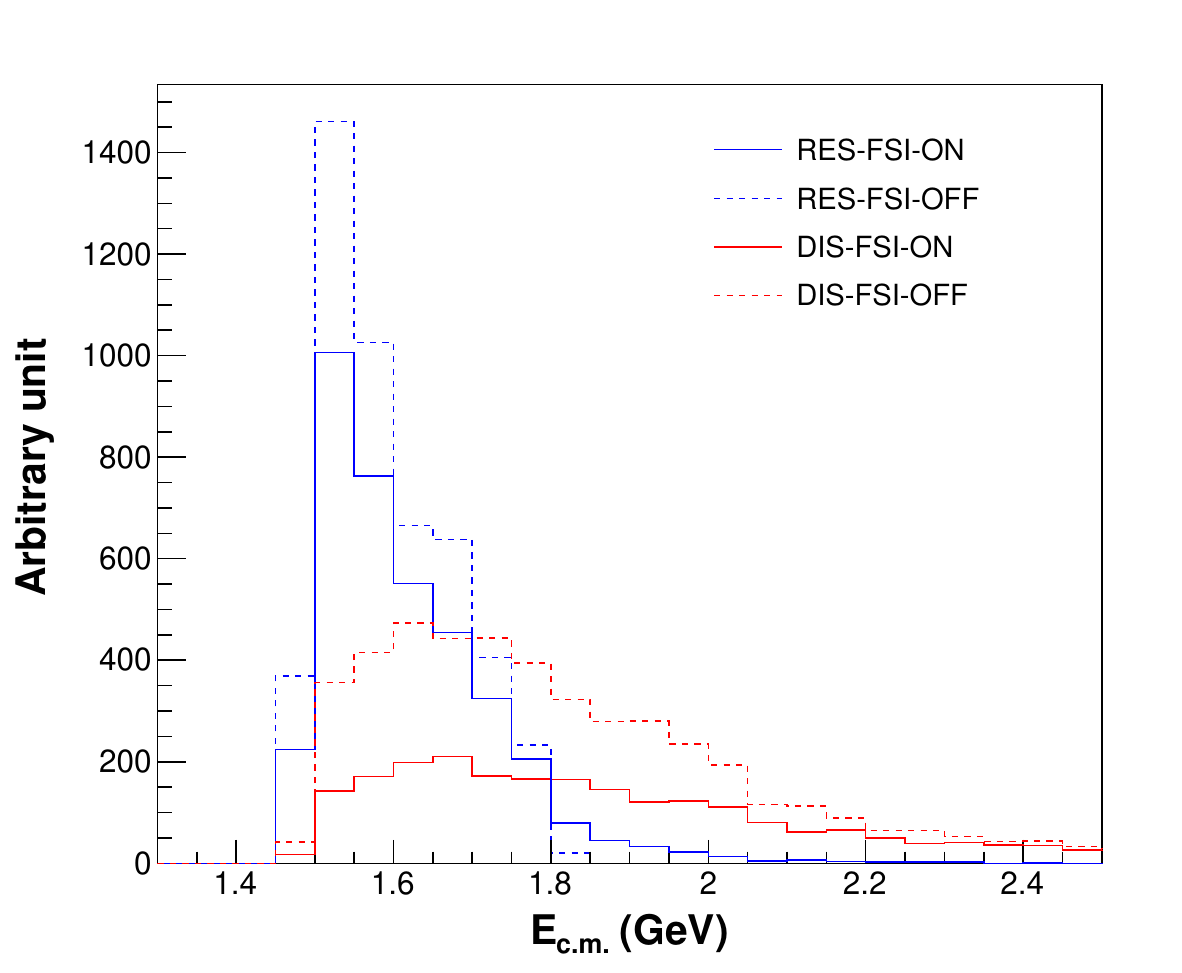}
    \caption{$E_{c.m.}$ distribution for FSI-ON/OFF with the contributions from RES and DIS to $\nu_{\mu}$CC1p1$\eta$ events.}
    \label{fig:3}
\end{figure}

A cut on the total energy in the c.m. frame can be used to select events with minimal FSI. $E_{c.m.}$ distribution for FSI-ON/OFF is shown in Fig. \ref{fig:3}, showing the contributions from RES and DIS events to the selection $\nu_{\mu}$CC1p1$\eta$. In the absence of FSI, the $E_{c.m.}$ equals the mass of the resonance state. The most contributions to $\eta$ production come from the $S_{11}(1535)$, $P_{11}(1710)$, $P_{11}(1440)$, and $P_{13}(1720)$ states. However, there are also small contributions from the $D_{13}(1520)$, $S_{11}(1650)$, $D_{13}(1700)$, $D_{15}(1675)$, and $F_{15}(1680)$ states. A cut of $1.44 \leq E_{c.m.} \,\text{(GeV)} \leq 1.72$ is applied to reduce the FSI effects from the event selection. The impact of the $E_{c.m.}$ cut on the $\theta_{c.m.}$ distribution is shown in Fig. \ref{fig:6} for both the hA and hN models, and is compared with the FSI-OFF distribution. The $\chi^2$/NDF is significantly reduced from 248.8/8 and 110.5/8 to 88.5/8 and 49.9/8 for the hA and hN models, respectively, when the $E_{c.m.}$ cut is applied. The higher $\chi^2$/NDF after applying the $E_{c.m.}$ cut could be due to contributions from $E_{c.m.}$ values that differ from the mass of the contributing resonance states. However, $E_{c.m.}$ cut can be useful to select no-FSI event when a specific resonance state is considered. 

\begin{figure*}
    \centering
    \includegraphics[width=19cm,height=6.5cm]{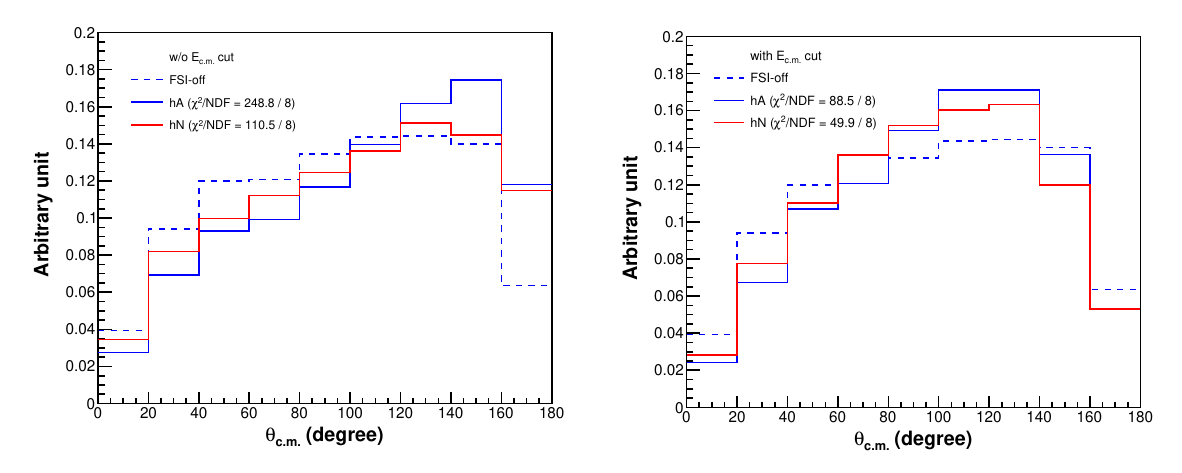}
    \caption{Area normalized distribution of $\theta_{c.m.}$ for FSI-ON/OFF using hA and hN model without (left) and with (right) $E_{c.m.}$ cut.}
    \label{fig:4}
\end{figure*}

\subsection{Dependency on nuclear initial state}

\begin{figure*}
    \centering
    \includegraphics[width=19cm,height=7cm]{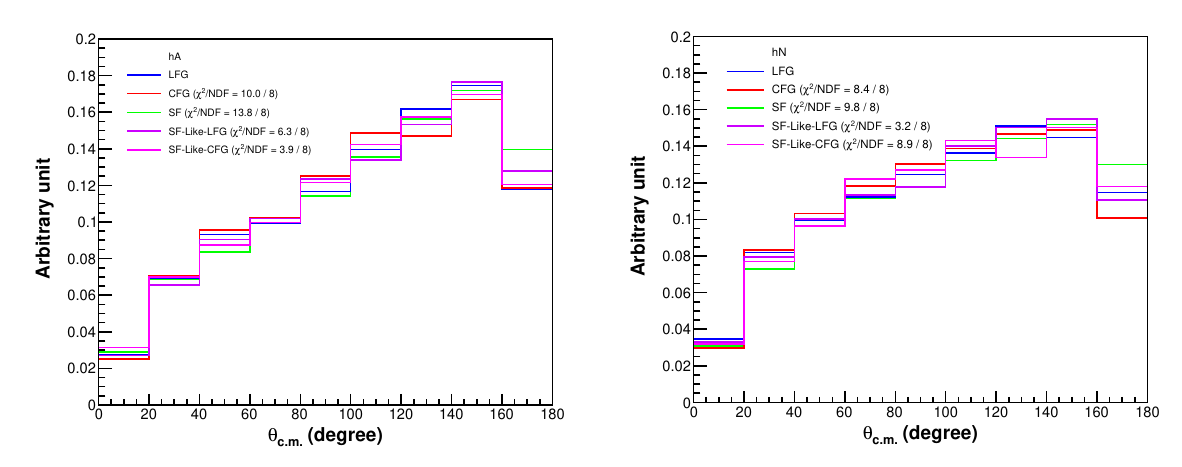}
    \caption{Area normalized distribution of $\theta_{c.m.}$ for different nuclear models using hA (left) and hN (right) models. }
    \label{fig:5}
\end{figure*}

\begin{figure*}
    \centering
    \includegraphics[width=19cm,height=6.5cm]{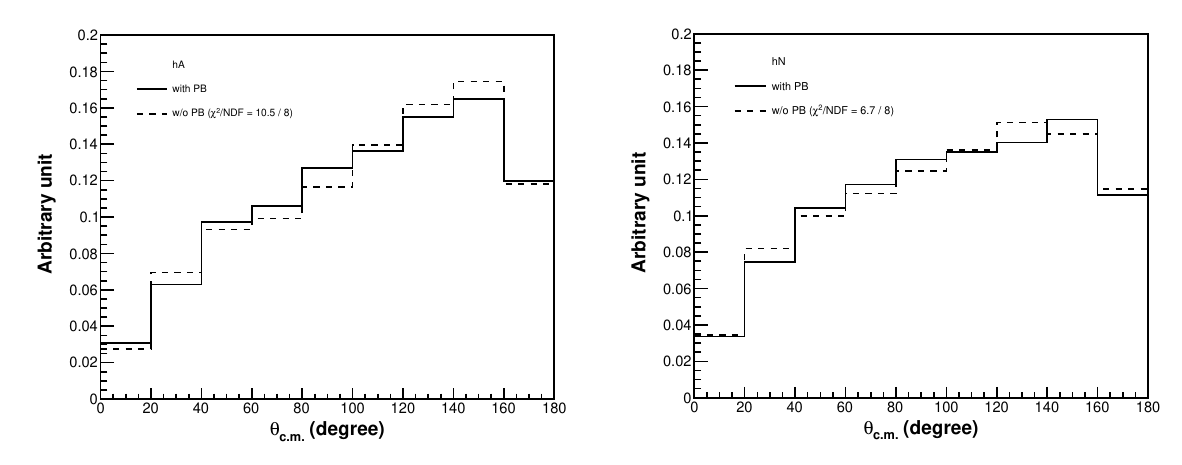}
    \caption{Area normalized distribution of $\theta_{c.m.}$ with and without Pauli blocking using hA (left) and hN (right) models. }
    \label{fig:6}
\end{figure*}

The figure \ref{fig:2} shows a minimal dependence of $\theta_{c.m.}$ on the initial state of the nucleus; however, a detailed impact of IS can be examined by reconstructing $\theta_{c.m.}$ for different IS effects, keeping other models fixed. $\theta_{c.m.}$ is calculated for different nuclear models shown in Fig. \ref{fig:5} using hA and hN FSI models. The distributions are close to each other with small $\chi^2$/NDF for hA. In the case of hN, there is a difference between the curves; however, $\chi^2$/NDF remains considerably small. This shows a negligible impact of nuclear models on $\theta_{c.m.}$. The next IS effect examined is the Pauli blocking (PB) \cite{Bodek:2021trq}. Pauli blocking restricts final-state nucleons from occupying already-filled energy states in the nucleus below Fermi state, which primarily suppresses interactions at low momentum transfer ($Q^2$). This reduces the phase space available for the final-state particle kinematics. The shape-comparison plot for $\theta_{c.m.}$ distribution with PB and without PB is shown in Fig. \ref{fig:6}. $\chi^2$/NDF is relatively small, 10.5/8 and 6.7/8 for hA and hN, respectively. However, the shape is different in the case of hN: the peak of the distribution is around 130$^\circ$ with PB and around 145$^\circ$ without PB. The difference in the case of hN could be due to the modeling of the hN model. The small $\chi^2$/NDF shows PB's low impact on $\theta_{c.m.}$.

\begin{figure*}
    \centering
    \includegraphics[width=19cm,height=6.5cm]{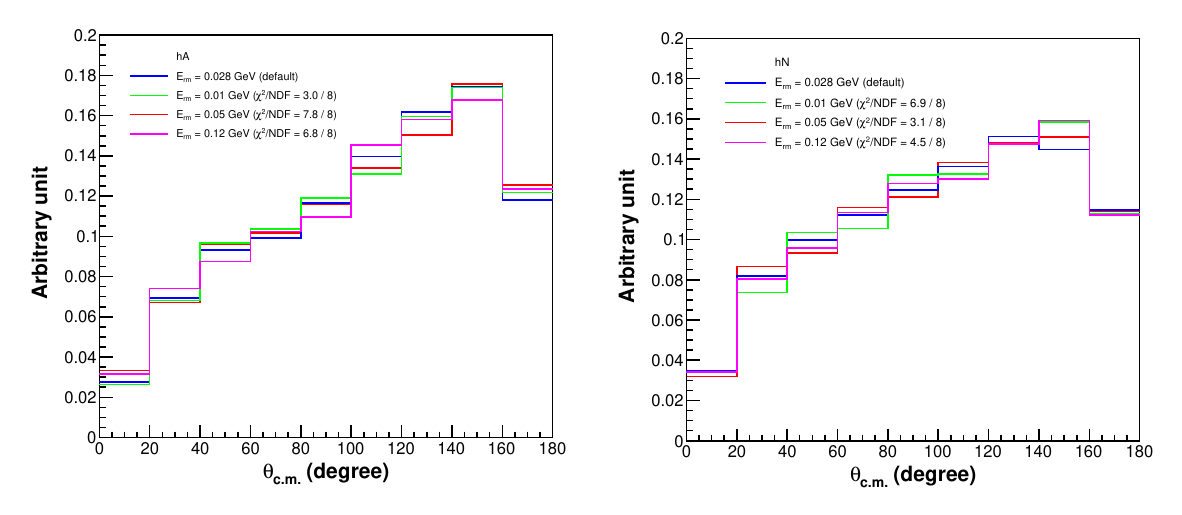}
    \caption{Area normalized distribution of $\theta_{c.m.}$ for different values of removal energy using hA (left) and hN (right) models. }
    \label{fig:7}
\end{figure*}

\begin{figure*}
    \centering
    \includegraphics[width=19cm,height=6.5cm]{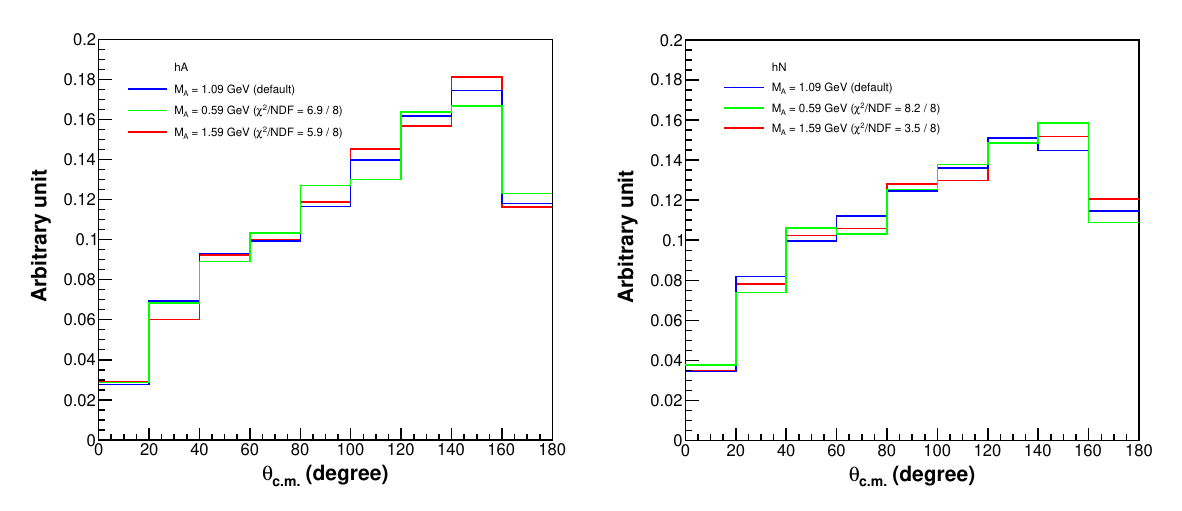}
    \caption{Area normalized distribution of $\theta_{c.m.}$ for different values of resonance axial mass using hA (left) and hN (right) models. }
    \label{fig:8}
\end{figure*}

\begin{figure}[!h]
    \centering
    \includegraphics[width=8.5cm,height=5.8cm]{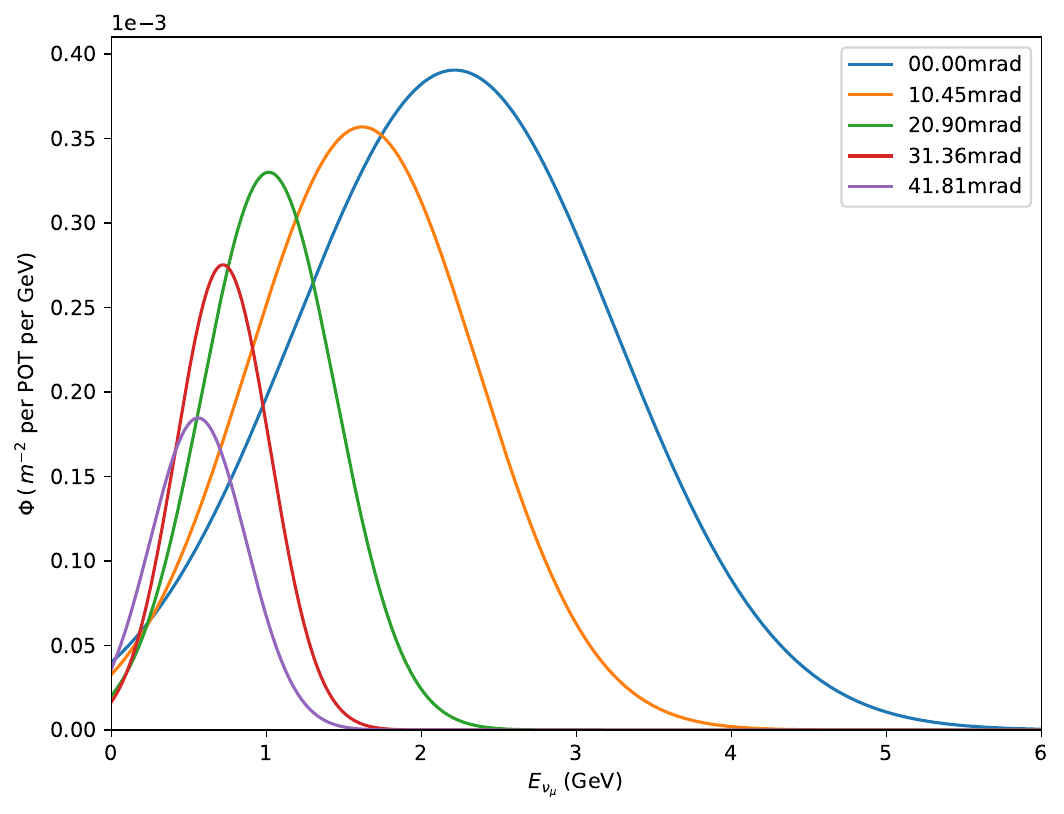}
    \caption{$\nu_{\mu}$ flux at different off-axis positions of DUNE-ND.}
    \label{prism_flux}
\end{figure}

\begin{figure*}
    \centering
    \includegraphics[width=19cm,height=7cm]{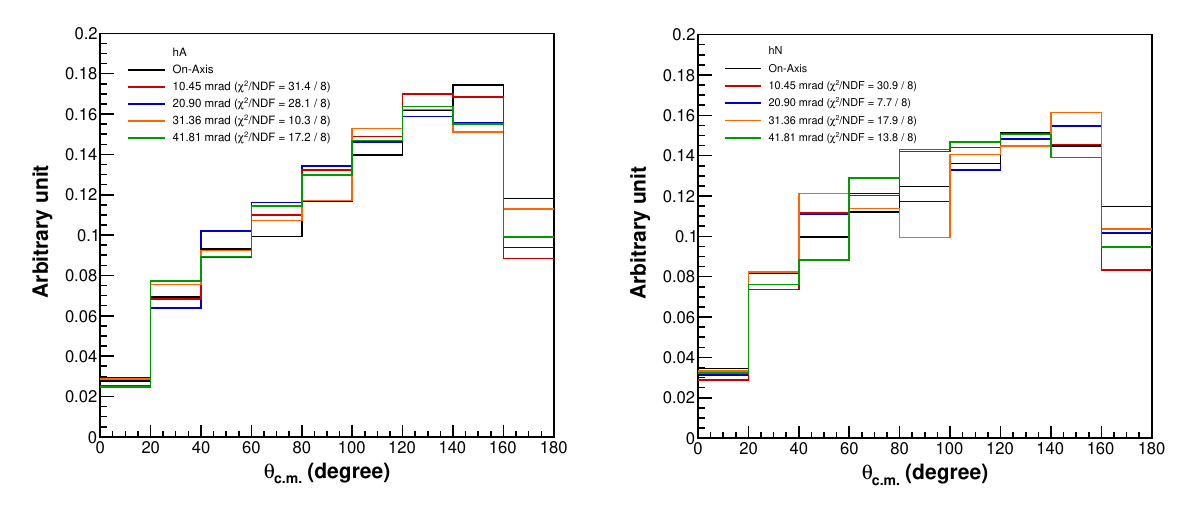}
    \caption{Area normalized distribution of $\theta_{c.m.}$ for different off-axis $\nu_{\mu}$ flux at DUNE-ND using hA (left) and hN (right) models.}
    \label{fig:9}
\end{figure*}

Another IS parameter that affects the neutrino interaction is the nucleon removal energy. It determines the energy needed to liberate a bound nucleon from the nucleus, and a higher removal energy reduces the visible energy in the final-state hadronic system. The default value of the removal energy for argon is taken as 28 MeV. To check the effect on $\theta_{c.m.}$, different values of removal energy, 10, 50, and 120 MeV, are considered. The comparison for different removal energies is shown in Fig. \ref{fig:7}. All the distributions remain close to each other with low $\chi^2$/NDF of 3.0/8, 7.8/8, and 6.8/8, respectively, for the hA model. In case of the hN model, the distributions deviate from each other; however, $\chi^2$/NDF remains small. This confirms the minimal impact of removal energy on $\theta_{c.m.}$. The cross section and kinematic distributions of resonance production can be influenced by the axial mass of the resonance ($M_A$) that modifies the $Q^2$ dependence of the axial form factor. The default value of $M_A$ is taken as 1.09 GeV. To investigate its impact, a low (0.59 GeV) and a high (1.59 GeV) unphysical value of $M_A$ are considered. There is a deviation between the distributions observed, shown in Fig. \ref{fig:8}; however, $\chi^2$/NDF remains small for both hA and hN models.

To explore the impact of incoming neutrino energy on $\theta_{c.m.}$, DUNE-PRISM \cite{DUNE:2021tad} is used for different off-axis $\nu_{\mu}$ fluxes. The proposed DUNE-ND has three components: LAr, GAr, and SAND. In the PRISM mechanism, LAr and GAr can move off-axis up to 30 m so that ND can measure different neutrino fluxes and be sensitive to all the interaction channels separately. The neutrino flux peak narrows and moves to lower energies as the detectors move off-axis, as shown in Fig. \ref{prism_flux}. $\theta_{c.m.}$ is calculated for different off-axis fluxes using the hA and hN models presented in Fig. \ref{fig:9}. The distributions show deviations from the on-axis positions; however, $\chi^2$/NDF remains comparatively small. Since the off-axis flux peak in the lower energy region below 2.5 GeV, the energy of the produced resonance states will be less, leading to a lower decay angle of $\eta$. But, large deviations are observed in the case of the hN model with $\chi^2$/NDF remaining small.

These results confirm the robustness of $\theta_{c.m.}$ against the IS effects for the hA model. However, an impact of IS on $\theta_{c.m.}$ is observed in the case of the hN model. The hN model is a full cascade model that considers the successive interactions of produced hadrons with any or all nucleons encountered along their path before leaving the nucleus \cite{electronsforneutrinos:2020tbf,Harewood:2019rzy}. This includes a dependence on local nuclear density in the cross-section calculation. The hN model also considers medium correction for both pions and nucleons \cite{Dytman:2021ohr}, including effects such as Pauli blocking and other in-medium effects to hadron propagation within the nucleus, leading to the IS dependence in the hN model.

\begin{figure}[!h]
    \centering
    \includegraphics[width=9.5cm,height=7cm]{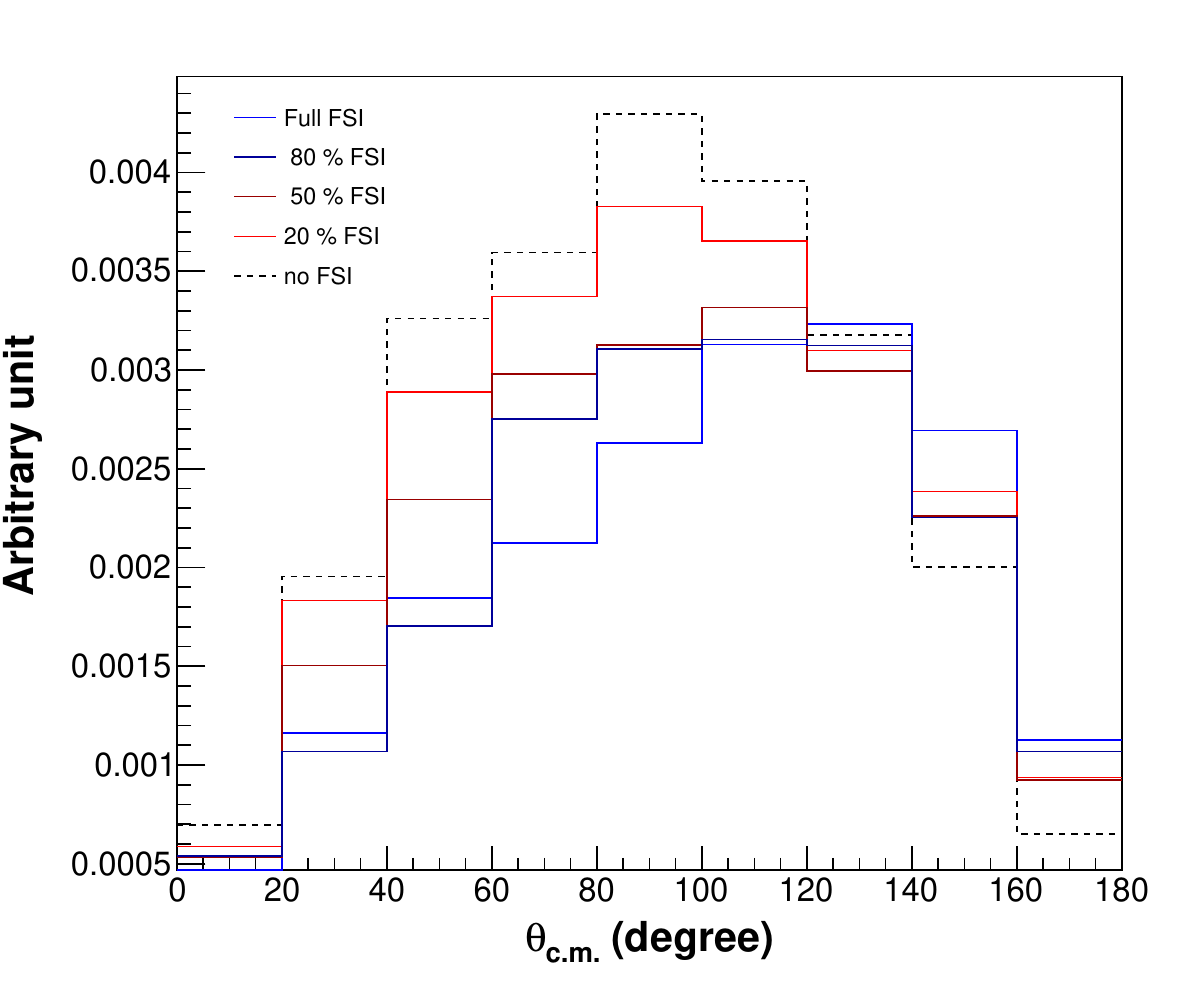}
    \caption{Cross section normalized distribution of $\theta_{c.m.}$ at different percentages of FSI.}
    \label{fig:10}
\end{figure}

\subsection{Variation of $\theta_{c.m.}$ with FSI fates}

\begin{figure*}
    \centering
    \includegraphics[width=19cm,height=7cm]{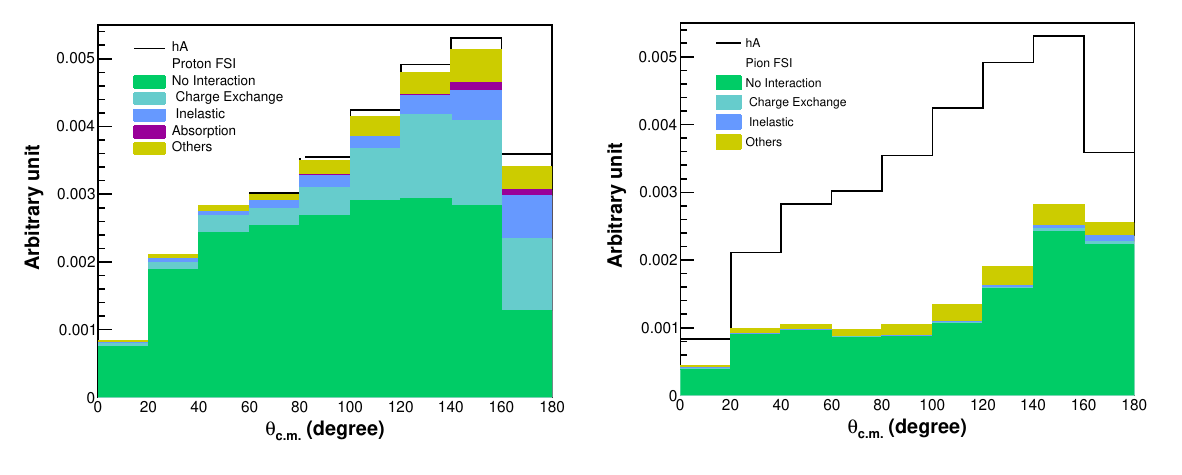}
    \caption{Cross section normalized distribution of $\theta_{c.m.}$ decomposed in FSI fates for proton (left) and pion (right).}
    \label{fig:11}
\end{figure*}

Since $\theta_{c.m.}$ is strongly influenced by the FSI effect with minimal dependence on IS, it becomes an efficient parameter to quantify FSI. The distribution of $\theta_{c.m.}$ is explored for different components of the hA model. Fig. \ref{fig:10} shows the variation in $\theta_{c.m.}$ with the percentages of FSI. FSI shifts the $\theta_{c.m.}$ towards the higher decay angle. The angular distribution is smeared out due to the rescattering of particles inside the nucleus, reducing the sharp features seen in the no-FSI case. The cross section tends to decrease in the region of forward angles ($\theta_{c.m.} < 50^\circ$ with increasing FSI. This indicates that FSI deflects the particles produced at a forward angle to larger angles. The decomposition of $\theta_{c.m.}$ distribution according to the different FSI fates is shown in Fig. \ref{fig:11} for both proton and pion. For the FSI effect on proton, $\theta_{c.m.}$ is mostly forward-peaked for the no-interaction case, indicating minimal deflection from the initial interaction. Forward peaking reflects the dominance of small momentum transfer in proton-nucleon collisions. When other fates like inelastic, absorption, and charge-exchange processes are included, the distribution broadens with increased cross section at large decay angles. In the case of pion FSI, the no-interaction has a broad distribution and peaks at high $\theta_{c.m.}$. There is an enhancement in cross section at intermediate and large angles ($\approx 60^\circ - 150^\circ$) when charge exchange and inelastic scattering are considered. This redistribution is a consequence of pion-nucleon interactions inside the nucleus, which alter the pion's trajectory through processes like charge exchange and multiple elastic scatterings. $\theta_{c.m.}$ can be useful to study a particular FSI fate and would be helpful to investigate the hypothesis of color transparency (CT) \cite{Gallmeister:2022gid,Niewczas:2019fro}.

\section{Conclusion \label{conclusion}}

This work explores the recently proposed novel variables: $\theta_{c.m.}$ and $E_{c.m.}$ \cite{Li:2025iiv} for the 1p1$\eta$ system in DUNE. The neutrino-induced $\eta$ production becomes crucial to study the higher resonances in neutrino interaction as well as for BSM studies. Future measurements on $\theta_{c.m.}$ would be helpful to understand the FSI in the resonance dominant energy regions and to reduce the uncertainty in neutrino cross section measurements. This work supports that $\theta_{c.m.}$ strongly depends on FSI and has a minimal dependence on IS for the 1p1$\eta$ in the final state. However, a dependence on IS is observed in the case of the hN model in contrast to the hA model. This dependence is due to the multiple scattering of produced hadrons within the nucleus and the consideration of medium corrections in the hN model. $\theta_{c.m.}$ also shows a dependence on the incoming neutrino DUNE-ND off-axis flux. The c.m. energy can be used to select high-purity no FSI events for a particular RES state. The c.m. variables are useful for studying different FSI fates, which can provide insight into the theoretical aspects of hadron rescattering within the nucleus and aid in exploring CT.

\section*{Acknowledgements} 
R K Pradhan acknowledges the DST-INSPIRE grant (2022/IF220293) for financial support. A Giri credited the grant support of the Department of Science and Technology (SR/MF/PS-01/2016-IITH/G).

\bibliography{references}

\begin{thebibliography}{37}%
\makeatletter
\providecommand \@ifxundefined [1]{%
 \@ifx{#1\undefined}
}%
\providecommand \@ifnum [1]{%
 \ifnum #1\expandafter \@firstoftwo
 \else \expandafter \@secondoftwo
 \fi
}%
\providecommand \@ifx [1]{%
 \ifx #1\expandafter \@firstoftwo
 \else \expandafter \@secondoftwo
 \fi
}%
\providecommand \natexlab [1]{#1}%
\providecommand \enquote  [1]{``#1''}%
\providecommand \bibnamefont  [1]{#1}%
\providecommand \bibfnamefont [1]{#1}%
\providecommand \citenamefont [1]{#1}%
\providecommand \href@noop [0]{\@secondoftwo}%
\providecommand \href [0]{\begingroup \@sanitize@url \@href}%
\providecommand \@href[1]{\@@startlink{#1}\@@href}%
\providecommand \@@href[1]{\endgroup#1\@@endlink}%
\providecommand \@sanitize@url [0]{\catcode `\\12\catcode `\$12\catcode
  `\&12\catcode `\#12\catcode `\^12\catcode `\_12\catcode `\%12\relax}%
\providecommand \@@startlink[1]{}%
\providecommand \@@endlink[0]{}%
\providecommand \url  [0]{\begingroup\@sanitize@url \@url }%
\providecommand \@url [1]{\endgroup\@href {#1}{\urlprefix }}%
\providecommand \urlprefix  [0]{URL }%
\providecommand \Eprint [0]{\href }%
\providecommand \doibase [0]{https://doi.org/}%
\providecommand \selectlanguage [0]{\@gobble}%
\providecommand \bibinfo  [0]{\@secondoftwo}%
\providecommand \bibfield  [0]{\@secondoftwo}%
\providecommand \translation [1]{[#1]}%
\providecommand \BibitemOpen [0]{}%
\providecommand \bibitemStop [0]{}%
\providecommand \bibitemNoStop [0]{.\EOS\space}%
\providecommand \EOS [0]{\spacefactor3000\relax}%
\providecommand \BibitemShut  [1]{\csname bibitem#1\endcsname}%
\let\auto@bib@innerbib\@empty
\bibitem [{\citenamefont {Abi}\ \emph {et~al.}(2020)\citenamefont {Abi} \emph
  {et~al.}}]{DUNE:2020ypp}%
  \BibitemOpen
  \bibfield  {author} {\bibinfo {author} {\bibfnamefont {B.}~\bibnamefont
  {Abi}} \emph {et~al.} (\bibinfo {collaboration} {DUNE}),\ }\bibfield  {title}
  {\bibinfo {title} {{Deep Underground Neutrino Experiment (DUNE), Far Detector
  Technical Design Report, Volume II: DUNE Physics}},\ }\href@noop {} {\
  (\bibinfo {year} {2020})},\ \Eprint {https://arxiv.org/abs/2002.03005}
  {arXiv:2002.03005 [hep-ex]} \BibitemShut {NoStop}%
\bibitem [{\citenamefont {Abe}\ \emph {et~al.}(2018)\citenamefont {Abe} \emph
  {et~al.}}]{Hyper-Kamiokande:2018ofw}%
  \BibitemOpen
  \bibfield  {author} {\bibinfo {author} {\bibfnamefont {K.}~\bibnamefont
  {Abe}} \emph {et~al.} (\bibinfo {collaboration} {Hyper-Kamiokande}),\
  }\bibfield  {title} {\bibinfo {title} {{Hyper-Kamiokande Design Report}},\
  }\href@noop {} {\  (\bibinfo {year} {2018})},\ \Eprint
  {https://arxiv.org/abs/1805.04163} {arXiv:1805.04163 [physics.ins-det]}
  \BibitemShut {NoStop}%
\bibitem [{\citenamefont {Balantekin}\ \emph {et~al.}(2022)\citenamefont
  {Balantekin} \emph {et~al.}}]{Balantekin:2022jrq}%
  \BibitemOpen
  \bibfield  {author} {\bibinfo {author} {\bibfnamefont {A.~B.}\ \bibnamefont
  {Balantekin}} \emph {et~al.},\ }\bibfield  {title} {\bibinfo {title}
  {{Snowmass Neutrino Frontier: Neutrino Interaction Cross Sections (NF06)
  Topical Group Report}},\ }\href@noop {} {\  (\bibinfo {year} {2022})},\
  \Eprint {https://arxiv.org/abs/2209.06872} {arXiv:2209.06872 [hep-ex]}
  \BibitemShut {NoStop}%
\bibitem [{\citenamefont {Aznauryan}\ \emph {et~al.}(2009)\citenamefont
  {Aznauryan} \emph {et~al.}}]{Aznauryan:2009da}%
  \BibitemOpen
  \bibfield  {author} {\bibinfo {author} {\bibfnamefont {I.}~\bibnamefont
  {Aznauryan}} \emph {et~al.},\ }\bibfield  {title} {\bibinfo {title} {{Theory
  Support for the Excited Baryon Program at the Jlab 12- GeV Upgrade}},\ }in\
  \href@noop {} {\emph {\bibinfo {booktitle} {{Electromagnetic N-N* Transition
  Form Factors Workshop}}}}\ (\bibinfo {year} {2009})\ \Eprint
  {https://arxiv.org/abs/0907.1901} {arXiv:0907.1901 [nucl-th]} \BibitemShut
  {NoStop}%
\bibitem [{\citenamefont {Rein}\ and\ \citenamefont
  {Sehgal}(1981)}]{Rein:1980wg}%
  \BibitemOpen
  \bibfield  {author} {\bibinfo {author} {\bibfnamefont {D.}~\bibnamefont
  {Rein}}\ and\ \bibinfo {author} {\bibfnamefont {L.~M.}\ \bibnamefont
  {Sehgal}},\ }\bibfield  {title} {\bibinfo {title} {{Neutrino Excitation of
  Baryon Resonances and Single Pion Production}},\ }\href
  {https://doi.org/10.1016/0003-4916(81)90242-6} {\bibfield  {journal}
  {\bibinfo  {journal} {Annals Phys.}\ }\textbf {\bibinfo {volume} {133}},\
  \bibinfo {pages} {79} (\bibinfo {year} {1981})}\BibitemShut {NoStop}%
\bibitem [{\citenamefont {Leitner}\ \emph {et~al.}(2009)\citenamefont
  {Leitner}, \citenamefont {Buss}, \citenamefont {Alvarez-Ruso},\ and\
  \citenamefont {Mosel}}]{Leitner:2008ue}%
  \BibitemOpen
  \bibfield  {author} {\bibinfo {author} {\bibfnamefont {T.}~\bibnamefont
  {Leitner}}, \bibinfo {author} {\bibfnamefont {O.}~\bibnamefont {Buss}},
  \bibinfo {author} {\bibfnamefont {L.}~\bibnamefont {Alvarez-Ruso}},\ and\
  \bibinfo {author} {\bibfnamefont {U.}~\bibnamefont {Mosel}},\ }\bibfield
  {title} {\bibinfo {title} {{Electron- and neutrino-nucleus scattering from
  the quasielastic to the resonance region}},\ }\href
  {https://doi.org/10.1103/PhysRevC.79.034601} {\bibfield  {journal} {\bibinfo
  {journal} {Phys. Rev. C}\ }\textbf {\bibinfo {volume} {79}},\ \bibinfo
  {pages} {034601} (\bibinfo {year} {2009})},\ \Eprint
  {https://arxiv.org/abs/0812.0587} {arXiv:0812.0587 [nucl-th]} \BibitemShut
  {NoStop}%
\bibitem [{\citenamefont {Abratenko}\ \emph
  {et~al.}(2024{\natexlab{a}})\citenamefont {Abratenko} \emph
  {et~al.}}]{MicroBooNE:2023ubu}%
  \BibitemOpen
  \bibfield  {author} {\bibinfo {author} {\bibfnamefont {P.}~\bibnamefont
  {Abratenko}} \emph {et~al.} (\bibinfo {collaboration} {MicroBooNE}),\
  }\bibfield  {title} {\bibinfo {title} {{First Measurement of
  \ensuremath{\eta} Meson Production in Neutrino Interactions on Argon with
  MicroBooNE}},\ }\href {https://doi.org/10.1103/PhysRevLett.132.151801}
  {\bibfield  {journal} {\bibinfo  {journal} {Phys. Rev. Lett.}\ }\textbf
  {\bibinfo {volume} {132}},\ \bibinfo {pages} {151801} (\bibinfo {year}
  {2024}{\natexlab{a}})},\ \Eprint {https://arxiv.org/abs/2305.16249}
  {arXiv:2305.16249 [hep-ex]} \BibitemShut {NoStop}%
\bibitem [{\citenamefont {Lalnuntluanga}\ and\ \citenamefont
  {Giri}(2023)}]{Lalnuntluanga:2023qkp}%
  \BibitemOpen
  \bibfield  {author} {\bibinfo {author} {\bibfnamefont {R.}~\bibnamefont
  {Lalnuntluanga}}\ and\ \bibinfo {author} {\bibfnamefont {A.}~\bibnamefont
  {Giri}},\ }\bibfield  {title} {\bibinfo {title} {{Quantifying the second
  resonance effect in neutrino-Argon interaction using DUNE Near Detector}},\
  }\href {https://doi.org/10.1016/j.physletb.2023.137717} {\bibfield  {journal}
  {\bibinfo  {journal} {Phys. Lett. B}\ }\textbf {\bibinfo {volume} {838}},\
  \bibinfo {pages} {137717} (\bibinfo {year} {2023})}\BibitemShut {NoStop}%
\bibitem [{\citenamefont {Adams}\ \emph {et~al.}(2020)\citenamefont {Adams}
  \emph {et~al.}}]{MicroBooNE:2019rgx}%
  \BibitemOpen
  \bibfield  {author} {\bibinfo {author} {\bibfnamefont {C.}~\bibnamefont
  {Adams}} \emph {et~al.} (\bibinfo {collaboration} {MicroBooNE}),\ }\bibfield
  {title} {\bibinfo {title} {{Reconstruction and Measurement of
  $\mathcal{O}$(100) MeV Energy Electromagnetic Activity from $\pi^0
  \rightarrow \gamma\gamma$ Decays in the MicroBooNE LArTPC}},\ }\href
  {https://doi.org/10.1088/1748-0221/15/02/P02007} {\bibfield  {journal}
  {\bibinfo  {journal} {JINST}\ }\textbf {\bibinfo {volume} {15}}\bibfield
  {number} {\bibinfo  {number} { (02)},\ \bibinfo {pages} {P02007}},\ }\Eprint
  {https://arxiv.org/abs/1910.02166} {arXiv:1910.02166 [hep-ex]} \BibitemShut
  {NoStop}%
\bibitem [{\citenamefont {Wittek}\ \emph {et~al.}(1989)\citenamefont {Wittek}
  \emph {et~al.}}]{BEBCWA59:1989ofp}%
  \BibitemOpen
  \bibfield  {author} {\bibinfo {author} {\bibfnamefont {W.}~\bibnamefont
  {Wittek}} \emph {et~al.} (\bibinfo {collaboration} {BEBC WA59}),\ }\bibfield
  {title} {\bibinfo {title} {{Production of $\rho^+$, $\rho^-$, $\rho^0$ (770),
  $\eta$ (550), $\omega(783)$ and F2 (1270) Mesons in Anti-nucleon Neon and
  Neutrino Neon Charged Current Interactions}},\ }\href
  {https://doi.org/10.1007/BF01557323} {\bibfield  {journal} {\bibinfo
  {journal} {Z. Phys. C}\ }\textbf {\bibinfo {volume} {44}},\ \bibinfo {pages}
  {175} (\bibinfo {year} {1989})}\BibitemShut {NoStop}%
\bibitem [{\citenamefont {Dytman}(2009)}]{Dytman:2009zz}%
  \BibitemOpen
  \bibfield  {author} {\bibinfo {author} {\bibfnamefont {S.}~\bibnamefont
  {Dytman}},\ }\bibfield  {title} {\bibinfo {title} {{Final state interactions
  in neutrino-nucleus experiments}},\ }\href@noop {} {\bibfield  {journal}
  {\bibinfo  {journal} {Acta Phys. Polon. B}\ }\textbf {\bibinfo {volume}
  {40}},\ \bibinfo {pages} {2445} (\bibinfo {year} {2009})}\BibitemShut
  {NoStop}%
\bibitem [{\citenamefont {Lalnuntluanga}\ \emph {et~al.}(2024)\citenamefont
  {Lalnuntluanga}, \citenamefont {Pradhan},\ and\ \citenamefont
  {Giri}}]{Lalnuntluanga:2024lti}%
  \BibitemOpen
  \bibfield  {author} {\bibinfo {author} {\bibfnamefont {R.}~\bibnamefont
  {Lalnuntluanga}}, \bibinfo {author} {\bibfnamefont {R.~K.}\ \bibnamefont
  {Pradhan}},\ and\ \bibinfo {author} {\bibfnamefont {A.}~\bibnamefont
  {Giri}},\ }\bibfield  {title} {\bibinfo {title} {{Probing neutrino-nucleus
  interaction in DUNE and MicroBooNE}},\ }\href
  {https://doi.org/10.1016/j.nuclphysb.2024.116703} {\bibfield  {journal}
  {\bibinfo  {journal} {Nucl. Phys. B}\ }\textbf {\bibinfo {volume} {1008}},\
  \bibinfo {pages} {116703} (\bibinfo {year} {2024})},\ \Eprint
  {https://arxiv.org/abs/2405.09994} {arXiv:2405.09994 [hep-ph]} \BibitemShut
  {NoStop}%
\bibitem [{\citenamefont {Lu}\ \emph {et~al.}(2016)\citenamefont {Lu},
  \citenamefont {Pickering}, \citenamefont {Dolan}, \citenamefont {Barr},
  \citenamefont {Coplowe}, \citenamefont {Uchida}, \citenamefont {Wark},
  \citenamefont {Wascko}, \citenamefont {Weber},\ and\ \citenamefont
  {Yuan}}]{Lu:2015tcr}%
  \BibitemOpen
  \bibfield  {author} {\bibinfo {author} {\bibfnamefont {X.~G.}\ \bibnamefont
  {Lu}}, \bibinfo {author} {\bibfnamefont {L.}~\bibnamefont {Pickering}},
  \bibinfo {author} {\bibfnamefont {S.}~\bibnamefont {Dolan}}, \bibinfo
  {author} {\bibfnamefont {G.}~\bibnamefont {Barr}}, \bibinfo {author}
  {\bibfnamefont {D.}~\bibnamefont {Coplowe}}, \bibinfo {author} {\bibfnamefont
  {Y.}~\bibnamefont {Uchida}}, \bibinfo {author} {\bibfnamefont
  {D.}~\bibnamefont {Wark}}, \bibinfo {author} {\bibfnamefont {M.~O.}\
  \bibnamefont {Wascko}}, \bibinfo {author} {\bibfnamefont {A.}~\bibnamefont
  {Weber}},\ and\ \bibinfo {author} {\bibfnamefont {T.}~\bibnamefont {Yuan}},\
  }\bibfield  {title} {\bibinfo {title} {{Measurement of nuclear effects in
  neutrino interactions with minimal dependence on neutrino energy}},\ }\href
  {https://doi.org/10.1103/PhysRevC.94.015503} {\bibfield  {journal} {\bibinfo
  {journal} {Phys. Rev. C}\ }\textbf {\bibinfo {volume} {94}},\ \bibinfo
  {pages} {015503} (\bibinfo {year} {2016})},\ \Eprint
  {https://arxiv.org/abs/1512.05748} {arXiv:1512.05748 [nucl-th]} \BibitemShut
  {NoStop}%
\bibitem [{\citenamefont {Abratenko}\ \emph
  {et~al.}(2024{\natexlab{b}})\citenamefont {Abratenko} \emph
  {et~al.}}]{MicroBooNE:2023krv}%
  \BibitemOpen
  \bibfield  {author} {\bibinfo {author} {\bibfnamefont {P.}~\bibnamefont
  {Abratenko}} \emph {et~al.} (\bibinfo {collaboration} {MicroBooNE}),\
  }\bibfield  {title} {\bibinfo {title} {{Measurement of nuclear effects in
  neutrino-argon interactions using generalized kinematic imbalance variables
  with the MicroBooNE detector}},\ }\href
  {https://doi.org/10.1103/PhysRevD.109.092007} {\bibfield  {journal} {\bibinfo
   {journal} {Phys. Rev. D}\ }\textbf {\bibinfo {volume} {109}},\ \bibinfo
  {pages} {092007} (\bibinfo {year} {2024}{\natexlab{b}})},\ \Eprint
  {https://arxiv.org/abs/2310.06082} {arXiv:2310.06082 [nucl-ex]} \BibitemShut
  {NoStop}%
\bibitem [{\citenamefont {Baudis}\ \emph {et~al.}(2024)\citenamefont {Baudis},
  \citenamefont {Dolan}, \citenamefont {Sgalaberna}, \citenamefont {Bolognesi},
  \citenamefont {Munteanu},\ and\ \citenamefont {Dieminger}}]{Baudis:2023tma}%
  \BibitemOpen
  \bibfield  {author} {\bibinfo {author} {\bibfnamefont {N.}~\bibnamefont
  {Baudis}}, \bibinfo {author} {\bibfnamefont {S.}~\bibnamefont {Dolan}},
  \bibinfo {author} {\bibfnamefont {D.}~\bibnamefont {Sgalaberna}}, \bibinfo
  {author} {\bibfnamefont {S.}~\bibnamefont {Bolognesi}}, \bibinfo {author}
  {\bibfnamefont {L.}~\bibnamefont {Munteanu}},\ and\ \bibinfo {author}
  {\bibfnamefont {T.}~\bibnamefont {Dieminger}},\ }\bibfield  {title} {\bibinfo
  {title} {{Longitudinal kinematic imbalances in neutrino and antineutrino
  interactions for improved measurements of neutrino energy and the axial
  vector form factor}},\ }\href {https://doi.org/10.1103/PhysRevD.110.032019}
  {\bibfield  {journal} {\bibinfo  {journal} {Phys. Rev. D}\ }\textbf {\bibinfo
  {volume} {110}},\ \bibinfo {pages} {032019} (\bibinfo {year} {2024})},\
  \Eprint {https://arxiv.org/abs/2310.15633} {arXiv:2310.15633 [hep-ph]}
  \BibitemShut {NoStop}%
\bibitem [{\citenamefont {Li}\ \emph {et~al.}(2024)\citenamefont {Li} \emph
  {et~al.}}]{GENIE:2024ufm}%
  \BibitemOpen
  \bibfield  {author} {\bibinfo {author} {\bibfnamefont {W.}~\bibnamefont {Li}}
  \emph {et~al.} (\bibinfo {collaboration} {GENIE}),\ }\bibfield  {title}
  {\bibinfo {title} {{First combined tuning on transverse kinematic imbalance
  data with and without pion production constraints}},\ }\href
  {https://doi.org/10.1103/PhysRevD.110.072016} {\bibfield  {journal} {\bibinfo
   {journal} {Phys. Rev. D}\ }\textbf {\bibinfo {volume} {110}},\ \bibinfo
  {pages} {072016} (\bibinfo {year} {2024})},\ \Eprint
  {https://arxiv.org/abs/2404.08510} {arXiv:2404.08510 [hep-ex]} \BibitemShut
  {NoStop}%
\bibitem [{\citenamefont {Yan}\ \emph {et~al.}(2024)\citenamefont {Yan},
  \citenamefont {Niewczas}, \citenamefont {Nikolakopoulos}, \citenamefont
  {Gonz\'alez-Jim\'enez}, \citenamefont {Jachowicz}, \citenamefont {Lu},
  \citenamefont {Sobczyk},\ and\ \citenamefont {Zheng}}]{Yan:2024kkg}%
  \BibitemOpen
  \bibfield  {author} {\bibinfo {author} {\bibfnamefont {Q.}~\bibnamefont
  {Yan}}, \bibinfo {author} {\bibfnamefont {K.}~\bibnamefont {Niewczas}},
  \bibinfo {author} {\bibfnamefont {A.}~\bibnamefont {Nikolakopoulos}},
  \bibinfo {author} {\bibfnamefont {R.}~\bibnamefont {Gonz\'alez-Jim\'enez}},
  \bibinfo {author} {\bibfnamefont {N.}~\bibnamefont {Jachowicz}}, \bibinfo
  {author} {\bibfnamefont {X.}~\bibnamefont {Lu}}, \bibinfo {author}
  {\bibfnamefont {J.}~\bibnamefont {Sobczyk}},\ and\ \bibinfo {author}
  {\bibfnamefont {Y.}~\bibnamefont {Zheng}},\ }\bibfield  {title} {\bibinfo
  {title} {{The Ghent Hybrid model in NuWro: a new neutrino single-pion
  production model in the GeV regime}},\ }\href
  {https://doi.org/10.1007/JHEP12(2024)141} {\bibfield  {journal} {\bibinfo
  {journal} {JHEP}\ }\textbf {\bibinfo {volume} {12}},\ \bibinfo {pages}
  {141}},\ \Eprint {https://arxiv.org/abs/2405.05212} {arXiv:2405.05212
  [hep-ph]} \BibitemShut {NoStop}%
\bibitem [{\citenamefont {Pradhan}\ \emph {et~al.}(2025)\citenamefont
  {Pradhan}, \citenamefont {Lalnuntluanga},\ and\ \citenamefont
  {Giri}}]{Pradhan:2024gqv}%
  \BibitemOpen
  \bibfield  {author} {\bibinfo {author} {\bibfnamefont {R.~K.}\ \bibnamefont
  {Pradhan}}, \bibinfo {author} {\bibfnamefont {R.}~\bibnamefont
  {Lalnuntluanga}},\ and\ \bibinfo {author} {\bibfnamefont {A.}~\bibnamefont
  {Giri}},\ }\bibfield  {title} {\bibinfo {title} {{Improving target neutron
  momentum reconstruction using MINER\ensuremath{\nu}A \ensuremath{\pi}0
  data}},\ }\href {https://doi.org/10.1103/PhysRevD.111.093002} {\bibfield
  {journal} {\bibinfo  {journal} {Phys. Rev. D}\ }\textbf {\bibinfo {volume}
  {111}},\ \bibinfo {pages} {093002} (\bibinfo {year} {2025})},\ \Eprint
  {https://arxiv.org/abs/2409.15913} {arXiv:2409.15913 [hep-ph]} \BibitemShut
  {NoStop}%
\bibitem [{\citenamefont {Li}(2025)}]{Li:2025iiv}%
  \BibitemOpen
  \bibfield  {author} {\bibinfo {author} {\bibfnamefont {W.}~\bibnamefont
  {Li}},\ }\bibfield  {title} {\bibinfo {title} {{Center-of-momentum variables
  in charged-current single-pion\textendash{}single-proton events}},\ }\href
  {https://doi.org/10.1103/PhysRevD.111.072010} {\bibfield  {journal} {\bibinfo
   {journal} {Phys. Rev. D}\ }\textbf {\bibinfo {volume} {111}},\ \bibinfo
  {pages} {072010} (\bibinfo {year} {2025})},\ \Eprint
  {https://arxiv.org/abs/2501.08984} {arXiv:2501.08984 [hep-ex]} \BibitemShut
  {NoStop}%
\bibitem [{\citenamefont {Workman}\ \emph {et~al.}(2022)\citenamefont {Workman}
  \emph {et~al.}}]{ParticleDataGroup:2022pth}%
  \BibitemOpen
  \bibfield  {author} {\bibinfo {author} {\bibfnamefont {R.~L.}\ \bibnamefont
  {Workman}} \emph {et~al.} (\bibinfo {collaboration} {Particle Data Group}),\
  }\bibfield  {title} {\bibinfo {title} {{Review of Particle Physics}},\ }\href
  {https://doi.org/10.1093/ptep/ptac097} {\bibfield  {journal} {\bibinfo
  {journal} {PTEP}\ }\textbf {\bibinfo {volume} {2022}},\ \bibinfo {pages}
  {083C01} (\bibinfo {year} {2022})}\BibitemShut {NoStop}%
\bibitem [{\citenamefont {Andreopoulos}\ \emph {et~al.}(2015)\citenamefont
  {Andreopoulos}, \citenamefont {Barry}, \citenamefont {Dytman}, \citenamefont
  {Gallagher}, \citenamefont {Golan}, \citenamefont {Hatcher}, \citenamefont
  {Perdue},\ and\ \citenamefont {Yarba}}]{Andreopoulos:2015wxa}%
  \BibitemOpen
  \bibfield  {author} {\bibinfo {author} {\bibfnamefont {C.}~\bibnamefont
  {Andreopoulos}}, \bibinfo {author} {\bibfnamefont {C.}~\bibnamefont {Barry}},
  \bibinfo {author} {\bibfnamefont {S.}~\bibnamefont {Dytman}}, \bibinfo
  {author} {\bibfnamefont {H.}~\bibnamefont {Gallagher}}, \bibinfo {author}
  {\bibfnamefont {T.}~\bibnamefont {Golan}}, \bibinfo {author} {\bibfnamefont
  {R.}~\bibnamefont {Hatcher}}, \bibinfo {author} {\bibfnamefont
  {G.}~\bibnamefont {Perdue}},\ and\ \bibinfo {author} {\bibfnamefont
  {J.}~\bibnamefont {Yarba}},\ }\bibfield  {title} {\bibinfo {title} {{The
  GENIE Neutrino Monte Carlo Generator: Physics and User Manual}},\ }\href@noop
  {} {\  (\bibinfo {year} {2015})},\ \Eprint {https://arxiv.org/abs/1510.05494}
  {arXiv:1510.05494 [hep-ph]} \BibitemShut {NoStop}%
\bibitem [{\citenamefont {Abi}\ \emph {et~al.}(2021)\citenamefont {Abi} \emph
  {et~al.}}]{DUNE:2021cuw}%
  \BibitemOpen
  \bibfield  {author} {\bibinfo {author} {\bibfnamefont {B.}~\bibnamefont
  {Abi}} \emph {et~al.} (\bibinfo {collaboration} {DUNE}),\ }\bibfield  {title}
  {\bibinfo {title} {{Experiment Simulation Configurations Approximating DUNE
  TDR}},\ }\href@noop {} {\  (\bibinfo {year} {2021})},\ \Eprint
  {https://arxiv.org/abs/2103.04797} {arXiv:2103.04797 [hep-ex]} \BibitemShut
  {NoStop}%
\bibitem [{\citenamefont {Bodek}\ and\ \citenamefont
  {Ritchie}(1981)}]{Bodek:1981wr}%
  \BibitemOpen
  \bibfield  {author} {\bibinfo {author} {\bibfnamefont {A.}~\bibnamefont
  {Bodek}}\ and\ \bibinfo {author} {\bibfnamefont {J.~L.}\ \bibnamefont
  {Ritchie}},\ }\bibfield  {title} {\bibinfo {title} {{Further Studies of Fermi
  Motion Effects in Lepton Scattering from Nuclear Targets}},\ }\href
  {https://doi.org/10.1103/PhysRevD.24.1400} {\bibfield  {journal} {\bibinfo
  {journal} {Phys. Rev. D}\ }\textbf {\bibinfo {volume} {24}},\ \bibinfo
  {pages} {1400} (\bibinfo {year} {1981})}\BibitemShut {NoStop}%
\bibitem [{\citenamefont {Egiyan}\ \emph {et~al.}(2006)\citenamefont {Egiyan}
  \emph {et~al.}}]{CLAS:2005ola}%
  \BibitemOpen
  \bibfield  {author} {\bibinfo {author} {\bibfnamefont {K.~S.}\ \bibnamefont
  {Egiyan}} \emph {et~al.} (\bibinfo {collaboration} {CLAS}),\ }\bibfield
  {title} {\bibinfo {title} {{Measurement of 2- and 3-nucleon short range
  correlation probabilities in nuclei}},\ }\href
  {https://doi.org/10.1103/PhysRevLett.96.082501} {\bibfield  {journal}
  {\bibinfo  {journal} {Phys. Rev. Lett.}\ }\textbf {\bibinfo {volume} {96}},\
  \bibinfo {pages} {082501} (\bibinfo {year} {2006})},\ \Eprint
  {https://arxiv.org/abs/nucl-ex/0508026} {arXiv:nucl-ex/0508026} \BibitemShut
  {NoStop}%
\bibitem [{\citenamefont {Benhar}\ \emph {et~al.}(1994)\citenamefont {Benhar},
  \citenamefont {Fabrocini}, \citenamefont {Fantoni},\ and\ \citenamefont
  {Sick}}]{Benhar:1994hw}%
  \BibitemOpen
  \bibfield  {author} {\bibinfo {author} {\bibfnamefont {O.}~\bibnamefont
  {Benhar}}, \bibinfo {author} {\bibfnamefont {A.}~\bibnamefont {Fabrocini}},
  \bibinfo {author} {\bibfnamefont {S.}~\bibnamefont {Fantoni}},\ and\ \bibinfo
  {author} {\bibfnamefont {I.}~\bibnamefont {Sick}},\ }\bibfield  {title}
  {\bibinfo {title} {{Spectral function of finite nuclei and scattering of GeV
  electrons}},\ }\href {https://doi.org/10.1016/0375-9474(94)90920-2}
  {\bibfield  {journal} {\bibinfo  {journal} {Nucl. Phys. A}\ }\textbf
  {\bibinfo {volume} {579}},\ \bibinfo {pages} {493} (\bibinfo {year}
  {1994})}\BibitemShut {NoStop}%
\bibitem [{\citenamefont {Gran}\ \emph {et~al.}(2013)\citenamefont {Gran},
  \citenamefont {Nieves}, \citenamefont {Sanchez},\ and\ \citenamefont
  {Vicente~Vacas}}]{Gran:2013kda}%
  \BibitemOpen
  \bibfield  {author} {\bibinfo {author} {\bibfnamefont {R.}~\bibnamefont
  {Gran}}, \bibinfo {author} {\bibfnamefont {J.}~\bibnamefont {Nieves}},
  \bibinfo {author} {\bibfnamefont {F.}~\bibnamefont {Sanchez}},\ and\ \bibinfo
  {author} {\bibfnamefont {M.~J.}\ \bibnamefont {Vicente~Vacas}},\ }\bibfield
  {title} {\bibinfo {title} {{Neutrino-nucleus quasi-elastic and 2p2h
  interactions up to 10 GeV}},\ }\href
  {https://doi.org/10.1103/PhysRevD.88.113007} {\bibfield  {journal} {\bibinfo
  {journal} {Phys. Rev. D}\ }\textbf {\bibinfo {volume} {88}},\ \bibinfo
  {pages} {113007} (\bibinfo {year} {2013})},\ \Eprint
  {https://arxiv.org/abs/1307.8105} {arXiv:1307.8105 [hep-ph]} \BibitemShut
  {NoStop}%
\bibitem [{\citenamefont {Nieves}\ \emph {et~al.}(2016)\citenamefont {Nieves},
  \citenamefont {Simo}, \citenamefont {S\'anchez},\ and\ \citenamefont
  {Vicente~Vacas}}]{Nieves:2016sma}%
  \BibitemOpen
  \bibfield  {author} {\bibinfo {author} {\bibfnamefont {J.}~\bibnamefont
  {Nieves}}, \bibinfo {author} {\bibfnamefont {I.~R.}\ \bibnamefont {Simo}},
  \bibinfo {author} {\bibfnamefont {F.}~\bibnamefont {S\'anchez}},\ and\
  \bibinfo {author} {\bibfnamefont {M.~J.}\ \bibnamefont {Vicente~Vacas}},\
  }\bibfield  {title} {\bibinfo {title} {{2p2h Excitations, MEC, Nucleon
  Correlations and Other Sources of QE-like Events}},\ }\href
  {https://doi.org/10.7566/JPSCP.12.010002} {\bibfield  {journal} {\bibinfo
  {journal} {JPS Conf. Proc.}\ }\textbf {\bibinfo {volume} {12}},\ \bibinfo
  {pages} {010002} (\bibinfo {year} {2016})}\BibitemShut {NoStop}%
\bibitem [{\citenamefont {Berger}\ and\ \citenamefont
  {Sehgal}(2007)}]{Berger:2007rq}%
  \BibitemOpen
  \bibfield  {author} {\bibinfo {author} {\bibfnamefont {C.}~\bibnamefont
  {Berger}}\ and\ \bibinfo {author} {\bibfnamefont {L.~M.}\ \bibnamefont
  {Sehgal}},\ }\bibfield  {title} {\bibinfo {title} {{Lepton mass effects in
  single pion production by neutrinos}},\ }\href
  {https://doi.org/10.1103/PhysRevD.76.113004} {\bibfield  {journal} {\bibinfo
  {journal} {Phys. Rev. D}\ }\textbf {\bibinfo {volume} {76}},\ \bibinfo
  {pages} {113004} (\bibinfo {year} {2007})},\ \Eprint
  {https://arxiv.org/abs/0709.4378} {arXiv:0709.4378 [hep-ph]} \BibitemShut
  {NoStop}%
\bibitem [{\citenamefont {Bodek}\ and\ \citenamefont
  {Yang}(2002)}]{Bodek:2002vp}%
  \BibitemOpen
  \bibfield  {author} {\bibinfo {author} {\bibfnamefont {A.}~\bibnamefont
  {Bodek}}\ and\ \bibinfo {author} {\bibfnamefont {U.~K.}\ \bibnamefont
  {Yang}},\ }\bibfield  {title} {\bibinfo {title} {{Modeling deep inelastic
  cross-sections in the few GeV region}},\ }\href
  {https://doi.org/10.1016/S0920-5632(02)01755-3} {\bibfield  {journal}
  {\bibinfo  {journal} {Nucl. Phys. B Proc. Suppl.}\ }\textbf {\bibinfo
  {volume} {112}},\ \bibinfo {pages} {70} (\bibinfo {year} {2002})},\ \Eprint
  {https://arxiv.org/abs/hep-ex/0203009} {arXiv:hep-ex/0203009} \BibitemShut
  {NoStop}%
\bibitem [{\citenamefont {Alvarez-Ruso}\ \emph {et~al.}(2021)\citenamefont
  {Alvarez-Ruso} \emph {et~al.}}]{GENIE:2021npt}%
  \BibitemOpen
  \bibfield  {author} {\bibinfo {author} {\bibfnamefont {L.}~\bibnamefont
  {Alvarez-Ruso}} \emph {et~al.} (\bibinfo {collaboration} {GENIE}),\
  }\bibfield  {title} {\bibinfo {title} {{Recent highlights from GENIE v3}},\
  }\href {https://doi.org/10.1140/epjs/s11734-021-00295-7} {\bibfield
  {journal} {\bibinfo  {journal} {Eur. Phys. J. ST}\ }\textbf {\bibinfo
  {volume} {230}},\ \bibinfo {pages} {4449} (\bibinfo {year} {2021})},\ \Eprint
  {https://arxiv.org/abs/2106.09381} {arXiv:2106.09381 [hep-ph]} \BibitemShut
  {NoStop}%
\bibitem [{\citenamefont {Bodek}(2021)}]{Bodek:2021trq}%
  \BibitemOpen
  \bibfield  {author} {\bibinfo {author} {\bibfnamefont {A.}~\bibnamefont
  {Bodek}},\ }\bibfield  {title} {\bibinfo {title} {{Pauli Blocking for a
  Relativistic Fermi Gas in Quasielastic Lepton Nucleus Scattering}},\
  }\href@noop {} {\  (\bibinfo {year} {2021})},\ \Eprint
  {https://arxiv.org/abs/2111.03631} {arXiv:2111.03631 [nucl-th]} \BibitemShut
  {NoStop}%
\bibitem [{\citenamefont {Hewes}\ \emph {et~al.}(2021)\citenamefont {Hewes}
  \emph {et~al.}}]{DUNE:2021tad}%
  \BibitemOpen
  \bibfield  {author} {\bibinfo {author} {\bibfnamefont {V.}~\bibnamefont
  {Hewes}} \emph {et~al.} (\bibinfo {collaboration} {DUNE}),\ }\bibfield
  {title} {\bibinfo {title} {{Deep Underground Neutrino Experiment (DUNE) Near
  Detector Conceptual Design Report}},\ }\href
  {https://doi.org/10.3390/instruments5040031} {\bibfield  {journal} {\bibinfo
  {journal} {Instruments}\ }\textbf {\bibinfo {volume} {5}},\ \bibinfo {pages}
  {31} (\bibinfo {year} {2021})},\ \Eprint {https://arxiv.org/abs/2103.13910}
  {arXiv:2103.13910 [physics.ins-det]} \BibitemShut {NoStop}%
\bibitem [{\citenamefont {Papadopoulou}\ \emph {et~al.}(2021)\citenamefont
  {Papadopoulou} \emph {et~al.}}]{electronsforneutrinos:2020tbf}%
  \BibitemOpen
  \bibfield  {author} {\bibinfo {author} {\bibfnamefont {A.}~\bibnamefont
  {Papadopoulou}} \emph {et~al.} (\bibinfo {collaboration} {electrons for
  neutrinos}),\ }\bibfield  {title} {\bibinfo {title} {{Inclusive Electron
  Scattering And The GENIE Neutrino Event Generator}},\ }\href
  {https://doi.org/10.1103/PhysRevD.103.113003} {\bibfield  {journal} {\bibinfo
   {journal} {Phys. Rev. D}\ }\textbf {\bibinfo {volume} {103}},\ \bibinfo
  {pages} {113003} (\bibinfo {year} {2021})},\ \Eprint
  {https://arxiv.org/abs/2009.07228} {arXiv:2009.07228 [nucl-th]} \BibitemShut
  {NoStop}%
\bibitem [{\citenamefont {Harewood}\ and\ \citenamefont
  {Gran}(2019)}]{Harewood:2019rzy}%
  \BibitemOpen
  \bibfield  {author} {\bibinfo {author} {\bibfnamefont {L.~A.}\ \bibnamefont
  {Harewood}}\ and\ \bibinfo {author} {\bibfnamefont {R.}~\bibnamefont
  {Gran}},\ }\bibfield  {title} {\bibinfo {title} {{Elastic hadron-nucleus
  scattering in neutrino-nucleus reactions and transverse kinematics
  measurements}},\ }\href@noop {} {\  (\bibinfo {year} {2019})},\ \Eprint
  {https://arxiv.org/abs/1906.10576} {arXiv:1906.10576 [hep-ex]} \BibitemShut
  {NoStop}%
\bibitem [{\citenamefont {Dytman}\ \emph {et~al.}(2021)\citenamefont {Dytman},
  \citenamefont {Hayato}, \citenamefont {Raboanary}, \citenamefont {Sobczyk},
  \citenamefont {Tena~Vidal},\ and\ \citenamefont
  {Vololoniaina}}]{Dytman:2021ohr}%
  \BibitemOpen
  \bibfield  {author} {\bibinfo {author} {\bibfnamefont {S.}~\bibnamefont
  {Dytman}}, \bibinfo {author} {\bibfnamefont {Y.}~\bibnamefont {Hayato}},
  \bibinfo {author} {\bibfnamefont {R.}~\bibnamefont {Raboanary}}, \bibinfo
  {author} {\bibfnamefont {J.~T.}\ \bibnamefont {Sobczyk}}, \bibinfo {author}
  {\bibfnamefont {J.}~\bibnamefont {Tena~Vidal}},\ and\ \bibinfo {author}
  {\bibfnamefont {N.}~\bibnamefont {Vololoniaina}},\ }\bibfield  {title}
  {\bibinfo {title} {{Comparison of validation methods of simulations for final
  state interactions in hadron production experiments}},\ }\href
  {https://doi.org/10.1103/PhysRevD.104.053006} {\bibfield  {journal} {\bibinfo
   {journal} {Phys. Rev. D}\ }\textbf {\bibinfo {volume} {104}},\ \bibinfo
  {pages} {053006} (\bibinfo {year} {2021})},\ \Eprint
  {https://arxiv.org/abs/2103.07535} {arXiv:2103.07535 [hep-ph]} \BibitemShut
  {NoStop}%
\bibitem [{\citenamefont {Gallmeister}\ and\ \citenamefont
  {Mosel}(2022)}]{Gallmeister:2022gid}%
  \BibitemOpen
  \bibfield  {author} {\bibinfo {author} {\bibfnamefont {K.}~\bibnamefont
  {Gallmeister}}\ and\ \bibinfo {author} {\bibfnamefont {U.}~\bibnamefont
  {Mosel}},\ }\bibfield  {title} {\bibinfo {title} {{Hadronization and Color
  Transparency}},\ }\href {https://doi.org/10.3390/physics4020029} {\bibfield
  {journal} {\bibinfo  {journal} {MDPI Physics}\ }\textbf {\bibinfo {volume}
  {4}},\ \bibinfo {pages} {440} (\bibinfo {year} {2022})},\ \Eprint
  {https://arxiv.org/abs/2202.12804} {arXiv:2202.12804 [nucl-th]} \BibitemShut
  {NoStop}%
\bibitem [{\citenamefont {Niewczas}\ and\ \citenamefont
  {Sobczyk}(2019)}]{Niewczas:2019fro}%
  \BibitemOpen
  \bibfield  {author} {\bibinfo {author} {\bibfnamefont {K.}~\bibnamefont
  {Niewczas}}\ and\ \bibinfo {author} {\bibfnamefont {J.~T.}\ \bibnamefont
  {Sobczyk}},\ }\bibfield  {title} {\bibinfo {title} {{Nuclear Transparency in
  Monte Carlo Neutrino Event Generators}},\ }\href
  {https://doi.org/10.1103/PhysRevC.100.015505} {\bibfield  {journal} {\bibinfo
   {journal} {Phys. Rev. C}\ }\textbf {\bibinfo {volume} {100}},\ \bibinfo
  {pages} {015505} (\bibinfo {year} {2019})},\ \Eprint
  {https://arxiv.org/abs/1902.05618} {arXiv:1902.05618 [hep-ex]} \BibitemShut
  {NoStop}%
\end{thebibliography}%

\end{document}